\documentclass[aps,preprint,amsmath,amssymb]{revtex4}
\usepackage{bm,color}
\usepackage{amsfonts}
\usepackage{graphicx}

\def\g5{\gamma_{5}}
\def\ga{\gamma}

\def\e{\epsilon}

\def\be{\begin{eqnarray}}
\def\ed{\end{eqnarray}}

\def\non{\nonumber}

\def\la{\langle}
\def\ra{\rangle}
\def\CH{H^{+}}
\def\tm{\tilde{m}}

\def\vpp{\vec{p}_P}
\def\vpv{\vec{p}_{V}}

\begin{document}
\title{ \bf  $B_{s,d}-\bar B_{s,d}$ mixings and $b\to q (\ga,\, \ell \bar \ell)$ decays
in private Higgs model}
\date{\today}

\author{ \bf Rachid Benbrik$^{1,2}$\footnote{Email:
rbenbrik@mail.ncku.edu.tw}, Chuan-Hung Chen$^{1,2}$\footnote{Email:
physchen@mail.ncku.edu.tw}, and Chao-Qiang
Geng$^{3}$\footnote{Email: geng@phys.nthu.edu.tw}
 }

\affiliation{ $^{1}$Department of Physics, National Cheng-Kung
University, Tainan 701, Taiwan \\
$^{2}$National Center for Theoretical Sciences, Hsinchu 300, Taiwan
\\
$^{3}$Department of Physics, National Tsing-Hua University, Hsin-Chu
, 300 Taiwan
 }

\begin{abstract}
We study the low energy phenomena induced by the lightest charged
Higgs in the private Higgs (PH) model, in which each quark flavor is
associated with a Higgs doublet. We show that the couplings of the
charged Higgs scalars  to fermions are fixed and the unknown
parameters are only the masses and mixing elements of the charged
Higgs scalars. As the charged Higgs masses satisfy with
$M_b<M_{c}\ll M_{s} \ll M_{d,u}$,  processes involving $B$-meson are
expected to be the ideal places to test the PH model. In particular,
we explore the constraints on the model from experimental data in
$B$ physics, such as the branching ratio (BR) and CP asymmetry (CPA)
of $B\to X_s \ga$, $B_{d,s}-\bar B_{d,s}$ mixings and the BR for
$B\to K^* \ell^{+} \ell^{-}$. We illustrate that the sign of the
Wilson coefficient for $B\to X_s \ga$ can be different from that in
the standard model,  while  this flipped sign  can be displayed by
the forward-backward asymmetry of $B\to V \ell^{+} \ell^{-}$ with
$V$ a vector meson. We  also demonstrate that $B_{d,s}-\bar B_{d,s}$
mixings and their time-dependent CPAs are negligible small and the
BR of $B\to K^* \ell^{+} \ell^{-}$ can have a more strict bound than
that of $B\to X_s \ga$.
\end{abstract}

\maketitle

\section{Introduction}

The masses of quarks and charged leptons are dictated by the Yukawa
sector in the standard model (SM) through the simple and elegant
Higgs mechanism, where the vacuum expectation value (VEV) of the
Higgs field, determined by massive gauge bosons, indicates the scale
of the electroweak symmetry breaking (EWSB) with $\langle H
\rangle=v/\sqrt{2}=246$ GeV. According to the data, there appear
mass hierarchies in the generations of charged fermions such as
$m_u\ll m_c \ll m_t$, $m_d\ll m_s \ll m_b$ and $m_e \ll m_\mu \ll
m_\tau$, while $m_t \gg m_b$ and $m_c \gg m_s$ but $m_u < m_d$
\cite{PDG08}. In the SM, due to the scale of the EWSB being fixed by
the VEV of the Higgs field, the mass hierarchies are ascribed to the
finetuning of the Yukawa couplings.

In the Cabibbo-Kobayashi-Maskawa (CKM) matrix~\cite{CKM}, defined by
$V_{CKM} =V^{U}_L V^{D^\dagger}_L$ with  the
unitary matrix $V^{U(D)}_{L}$ for diagonalizing the quark mass matrices, it is
known that the off-diagonal elements denoted by $(V_{CKM})_{i\neq
j}$ are suppressed by the Wolfenstein parameter $\lambda$
\cite{Wolfenstein}. If the effects of $\lambda$ are turned off,  one
immediately finds $V^{U}_{L}= V^{D}_L$. In other words, small
elements of $(V_{CKM})_{i\neq j}$ imply that the structures of the
Yuwaka matrices for up and down type quarks should be close to each
other. However, based on the above discussion, the similarity of the
mass structures is not respected by the data. Plausibly, we need to
extend the Yukawa sector to explain the mass hierarchies.


In order to evade the drawback of the finetuned Yukawa couplings, a
new type of solutions to the mass hierarchy is recently  proposed in
Refs.~\cite{PZ1,PZ2}, in which the authors extend one Higgs doublet
in the SM to multi-Higgs doublets with each gauge singlet
right-handed fermion  associated with one Higgs doublet.
 Hereafter, the model is called as the private Higgs
(PH) model \cite{PZ1}. The philosophy of solving the mass hierarchies
in generations is now to utilize the hierarchy of VEVs of scalar
fields instead of the hierarchy of the Yukawa couplings.
Although  many
new neutral and charged scalar bosons are introduced in the PH
model,  most of the effects are suppressed by the heavy
masses.
In addition, the PH model provides the candidate of dark matter. The
detailed study could be referred to Ref.~\cite{Jackson}.

Since top and bottom quarks are the first two heaviest fermions, the
dominant new effects are expected to be associated with the Higgs
doublets, denoted by $\Phi_{t,b}$, respectively. Since $m_t \gg m_b$
implies $\langle \Phi_{t}\rangle \gg \langle \Phi_{b}\rangle$,
$\Phi_b$ gives the dominant new physical effects if we take $\Phi_t$
as the SM Higgs. Accordingly, we anticipate that the B-meson system
could be the good environment to probe the special character in the
PH model. In this paper, we study the effects of the private charged
Higgs bosons on the rare flavor changing neutral current (FCNC)
processes, such as $B_{q}-\bar B_{q}$ mixings and $b \to q(\gamma,
\ell^+ \ell^-)$ decays with q=s, d. These processes are expected to
be
sensitive to the charged Higgs sector.

The paper is organized as follows. In Sec.~\ref{sec:PHM}, we briefly
summarize the PH model. In Sec.~\ref{sec:phe}, we study the
contributions of the  charged Higgs scalars on $B_q-\bar B_q$
mixings, $B\to X_s\ga$, $B_q\to \ell^{+} \ell^{-}$ and $B \to (P,\,
V)\ell^{+} \ell^{-}$ decays. The numerical results and their
discussions are given in Sec.~\ref{sec:num}. Finally, we present the
summary in Sec.~\ref{sec:summary}.


\section{Charged Higgses in private Higgs model} \label{sec:PHM}

To examine the charged-Higgs effects in the PH model, we first
review the model proposed in Ref.~\cite{PZ1}. In the model, as the
hierarchy of the scalar VEVs is used to understand the fermion masses
instead of  the arbitrary Yukawa couplings, the SM with one Higgs
doublet is extended to include six Higgs doublets so that each Higgs
doublet can only couple to one flavor with imposing a set of six
$Z_{2}$ discrete symmetries. In addition,  six gauge singlet real
scalars are introduced to achieve the spontaneous symmetry
breakings. For simplicity, we will take only one singlet scalar
field $S$ in our discussion. The six-singlet case can be easily
accommodated but our results on FCNCs remain the same. Under the
discrete symmetries, the transformations for the flavor and the
scalars are set to be
 \be
 f_{R} \to -f_{R}\,,\ \ \ \Phi_{f}\to -\Phi_{f}\,, \ \ \ S\to -S\,,
 \ed
where $f$ denotes the possible flavor of the quark, $\Phi_{f}$ is
the associated Higgs doublet scalar and S is the gauge singlet
scalar. Since the left-handed quark belongs to the $SU(2)$ doublet
of two flavors, we require that it is invariant under the discrete
transformations. Accordingly, the related scalar interactions with
the electroweak gauge and  $Z_2$ discrete symmetries are given by
\cite{PZ1}
 \be
{\cal L}&=& \partial_{\mu} S \partial^{\mu} S -\frac{\lambda_s}{4}
\left(S^2-\frac{v^2_x}{2} \right)^2+\sum_{f} \left[
(D_{\mu}\Phi_{f})^{\dagger} (D^{\mu}\Phi_{f}) -\frac{1}{2} M^2_{f}
\Phi^{\dagger}_{f}\Phi_{f}-\lambda_{f}
\left( \Phi^{\dagger}_{f} \Phi_{f}\right)^2  \right. \non \\
&&\left. +g_{sf} S^2 \Phi^{\dagger}_{f} \Phi_{f}\right] +
\sum_{f\neq f'} \left[ \frac{\ga_{ff'}}{\sqrt{2}}v_s S
\Phi^{\dagger}_{f} \Phi_{f'}+ a_{ff'} \Phi^{\dagger}_{f} \Phi_{f'}
\Phi^{\dagger}_{f} \Phi_{f'} + b_{ff'} \Phi^{\dagger}_{f} \Phi_{f}
\Phi^{\dagger}_{f'} \Phi_{f'} \right.\non \\
&& \left. + c_{ff'} \Phi^{\dagger}_{f} \Phi_{f'}\Phi^{\dagger}_{f'}
\Phi_{f} \right]-{\cal L}_{Y}\,,
 \ed
where $D_{\mu}= i\partial_{\mu} -g \frac{\vec{\tau}}{2}\cdot
\vec{W}_{\mu} -g' \frac{Y}{2}\, B_{\mu}$ is the covariant
derivative, $M_f$ is the mass of $\Phi_{f}$, $v_s$ is the VEV of S,
$v_{x}$ is a free parameter, the values of $\ga_{ff'}$, $a_{ff'}$,
$b_{ff'}$ and $c_{ff'}$ are regarded as the same order of magnitude,
and ${\cal L}_{Y}$ stands for the Yukawa sector to be given. Since
the top quark is the heaviest quark with its mass  close to the EWSB
scale, it is natural to take $\Phi_{t}$ as the Higgs doublet in the
SM. Therefore, to develop a nonzero VEV of $\Phi_{t}$ to have the
EWSB spontaneously, the condition of $M^2_{t}/2 < g_{st} v^2_{s}$
should be satisfied. Consequently, the relevant scalar potential
with the leading terms is given by
 \be
 V^{LT}&=&\frac{\lambda_s}{4}
\left(S^2-\frac{v^2_x}{2} \right)^2 + \lambda_t\left(
\Phi^{\dagger}_{t} \Phi_{t}\right)^2- g_{st}S^2 \Phi^{\dagger}_{t}
\Phi_{t}\,. \label{eq:vlo}
 \ed
By minimizing Eq.~(\ref{eq:vlo}), the VEVs of S and $\Phi_t$ are
obtained as
 \be
 \la S\ra^2 &\equiv& \frac{v^2_{s}}{2}=\frac{1}{2} \frac{\lambda_s \lambda_t}{\lambda_s
 \lambda_t-g^2_{st}} v^2_{x}\,, \non \\
 \la \Phi^0_{t} \ra^2 & \equiv& \frac{v^2_{t}}{2}=
 \frac{g_{st}}{2\lambda_t}v^2_{s}\,.
 \ed

We now discuss how to get the small VEVs for $\Phi_{f\neq t}$.
Unlike the case for $\Phi_t$, we need to adopt the condition $M_{f}
> \sqrt{g_{sf}} v_s$ for $f\neq t$.
The relevant subleading scalar potential for $f\neq t$ is
 \be
 V^{SLT}&=& \sum_{f\neq t} \left[ \frac{1}{2} M^{2}_{f} \Phi^{\dagger}_{f} \Phi_{f} -
 \left(\frac{\gamma_{tf}}{\sqrt{2}}v_s S \Phi^{\dagger}_{t}
 \Phi_{f}+h.c\right) \right]\,. \label{eq:vslo}
 \ed
We note that although the coefficients of $a_{tf'}$, $b_{tf'}$ and
$c_{tf'}$ are similar to $\ga_{tf'}$ in magnitude, their effects are
sub-subleading and negligible due to the associated VEVs of scalar
fields being much less than $v_s$. Similarly, by minimizing
Eq.~(\ref{eq:vslo}) the VEV of $\Phi^{0}_{f\neq t}$ is given by
\cite{PZ1}
 \be
 \la \Phi^{0}_{f} \ra= \ga_{tf}\frac{v_t}{\sqrt{2}}
 \frac{v^2_s}{2M^2_{f}}\,. \label{eq:vev}
 \ed
Clearly, if we set $\gamma_{tf}$  to be the same order of magnitude
for a different $f$, the hierarchy of VEVs could be obtained by
controlling $M_{f}$, i.e., the heavier $M_{f}$ is, the smaller $\la
\Phi^{0}_{f} \ra$ will be for $f\neq t$.

After introducing the strategy to obtain the EWSB spontaneously as
well as the small VEVs of the scalar fields with $f\neq t$, we can
proceed to investigate the characters of the charged Higgs scalars
in the PH model. In terms of $SU(2)_{L}\times U(1)_{Y}$ gauge
symmetries, the Yukawa sector is given by
 \be
 {\cal L }_{Y} &=& - \bar Q'_L Y_{D}  {\bf \Phi_{D}} d'_R - \bar Q'_L Y_{U} {\bf
 \tilde \Phi_{U}}
 u'_R + h.c. \,, \label{eq:Yukawa}
 \ed
where $Q'_{L}=(u', d')_{L}$ and $q'_R$ denote the doublet and
singlet of $SU(2)_{L}$, respectively, and $Y_{D(U)}$ is the $3\times
3$ Yukawa matrix for down (up) type quarks. In the flavor space,
${\bf \Phi_{D, U}}$ are also  $3\times 3$ matrices, given by
 \be
 {\bf \Phi_{D}}&=& \left(
                  \begin{array}{ccc}
                    \Phi_d & 0 & 0 \\
                    0 & \Phi_s & 0 \\
                    0 &0 & \Phi_b \\
                  \end{array}
                \right)\,,\ \ \
 {\bf \tilde \Phi_{U}}= \left(
                  \begin{array}{ccc}
                    \tilde \Phi_u & 0 & 0 \\
                    0 & \tilde \Phi_c & 0 \\
                    0 &0 & \tilde \Phi_t \\
                  \end{array}
                \right)\,, \label{eq:Phi_F}
 \ed
where $\Phi^T_{D}=(\phi^{+}, \phi^{0})_{D}$ and $\tilde
\Phi_{U}=i\tau_2 \Phi^{*}_{U}$  are
the Higgs doublets of $SU(2)_L$, which couple to
$D=(d,s,b)$ and $U=(u,c,t)$,
respectively. After the EWSB with the  shifted scalar fields
 \be
\phi^{0}_{F}= \frac{1}{\sqrt{2}}\left( v_{F}+ H_{F} +i A_F \right),\
(F=D\,,\ U)\,,\non
  \ed
the mass terms of quarks in the Yukawa sector are developed to be
 \be
 {\cal L}_{mass}&=& -\bar d'_{L} Y_D \frac{{\bf V_D}}{\sqrt{2}} d'_{R}
 - \bar u'_{L} Y_U \frac{{\bf V_U}}{\sqrt{2}} u'_{R}+h.c
 \label{eq:mass_W}
 \ed
with
 \be
 {\bf V}_{D(U)}=\left(
               \begin{array}{ccc}
                 v_{d(u)} & 0 & 0 \\
                 0 & v_{s(c)} & 0 \\
                 0 & 0 & v_{b(t)} \\
               \end{array}
             \right)\,. \label{eq:vev_F}
 \ed
To avoid the large FCNCs at tree level, we adopt the
Yukawa matrices in
Ref.~\cite{PZ1}, given by
 \be
  Y^{Q}_{ij}= \lambda_Q \delta_{ij} + \epsilon^{Q}_{ij}\,,
  \label{eq:YQ}
 \ed
where $Q=U$ and $D$,
$\lambda_{Q}\sim O(1)$ and $\epsilon^{Q}\ll 1$. By combining with
$v_{d(u)}\ll v_{s(c)}\ll v_{b(t)}$, the quark mass matrices  can be
simplified as
 \be
 M_{D}= \left(
             \begin{array}{ccc}
          ({\bf m}_d)_{2\times 2}  & | & {\bm \epsilon_d}_{2\times 1}
                       \\
             --- & | & --\\
            {\bf 0}_{ 1\times 2}  & | & m_b  \\
             \end{array}
           \right)\,,\ \ \
 M_{U}= \left(
             \begin{array}{ccc}
          ({\bf m}_u)_{2\times 2}  & | & {\bm \epsilon_u}_{2\times 1}
                       \\
             --- & | & --\\
            {\bf 0}_{ 1\times 2}  & | & m_t  \\
             \end{array}
           \right) \label{eq:massMa}
 \ed
where ${\rm dia}({\bf m}_{d(u)}) = \lambda_{D(U)} ( v_{d(u)},
v_{s(u)})/\sqrt{2}$ correspond to the light quarks and
$m_{b(t)}=\lambda_{D(U)} v_{b(t)}/\sqrt{2}$  the first two heaviest quarks,
${\bm \epsilon_{d(u)}}_{11}=\epsilon^{D(U)}_{13} v_{b(t)}/\sqrt{2}$
and ${\bm \epsilon_{d(u)}}_{21}=\epsilon^{D(U)}_{23}
v_{b(t)}/\sqrt{2}$. To get the physical states, we use $V^{D}_{L,
R}$ and $V^{U}_{L,R}$ to diagonalize the mass matrices,
i.e. $M^{\rm dia}_{U} = V^U_{L} M_{U} V^{U\dagger}_R$ and $M^{\rm
dia}_{D} = V^D_{L} M_{D} V^{D\dagger}_R$. The individual informations
on $V^{Q}_L$ and $V^{Q}_R$ can be obtained by
 \be
 M^{\rm dia}_{Q}M^{\rm dia^\dagger}_{Q}   &=& V^Q_{L}
 M_{Q} M^\dagger_{Q} V^{Q\dagger}_L \,,\non \\
 M^{\rm dia^\dagger}_{Q} M^{\rm dia}_{Q}  &=& V^{Q}_{R}
 M^\dagger_{Q} M_{Q} V^{Q\dagger}_R\,,
\label{eq:M_2}
 \ed
 respectively, where
 \be
 M_{Q} M^\dagger_{Q} = \left(
             \begin{array}{ccc}
             {\bf m}_q {\bf m}^\dagger_q +{\bm \epsilon}_q {\bm\epsilon
}^\dagger_q & | & {\bm\epsilon}_q m_H
             \\
             --- & | & --\\
             m_H {\bm\epsilon}^\dagger_{q}  & | & m^2_H \\
             \end{array}
           \right)\,, \ \ \
 M_{Q}^\dagger M_{Q} = \left(
             \begin{array}{ccc}
          {\bf m}^\dagger_q {\bf m}_q & | & {\bf m}^\dagger_q {\bm\epsilon}_{q}
             \\
             --- & | & --\\
  {\bm \epsilon}^\dagger_{q} {\bf m}_q & | & m^2_H
  +{\bm\epsilon}^\dagger_{q}{\bm\epsilon}_{q}  \\
             \end{array}
           \right) \label{eq:msquare}
 \ed
 with $m_H=(m_b, m_t)$. Due to $m_H\gg \epsilon_q, m_q$,  it is
 a good approximation to take $V^{Q}_{R(L)}\approx 1 +
\Delta^{Q}_{R(L)}$. Furthermore, from Eq.~(\ref{eq:msquare}), one
 observes that the off-diagonal elements of $M_{Q}M^{\dagger}_{Q}$
are much larger than those of $M^{\dagger}_{Q}M_{Q}$ and thus,
$\Delta^{Q}_{L}\sim O(\epsilon_q/m_H)$ and
$\Delta^{Q}_{R}\sim O(m_q\epsilon_q/m^2_H)$. As a result, at the leading
order approximation the right-handed unitary matrices could be taken
as identity matrices. Consequently,
we obtain
 \be
 \left(\Delta^{Q}_{L}\right)_{i3} &=&  -\left(\Delta^{Q^*}_L\right)_{3i}\approx -
 \frac{{\bm\epsilon_q}_{i1}}{m_H}=
 -\frac{\epsilon^Q_{i3}}{\lambda_Q}\,. \label{eq:delta}
 \ed
It is clear that
the induced FCNCs at tree level due to the
Yukawa terms are
suppressed by $\epsilon^Q_{i3}/\lambda_D$. Although we cannot get a
simple relation for $(\Delta^Q_{L})_{ij}$ with $i,j<3$,
the FCNCs,
involving the first two generations at
tree level, will be suppressed by the heavy masses of $\phi_{d, s,
u}$. The detailed analysis on the neutral Higgs exchange can be found in
Ref.~\cite{PZ1}.


In order to demonstrate that the neutral Higgs mediated FCNC effects will
not impose a further serious constraint on the parameters for the charged
Higgs, below we give an explicit discussion  on the $B_q-\bar B_q$
mixing. According to Eq.~(\ref{eq:Yukawa}), the relevant
Yukawa terms are given by
 \be
 {\cal L}_{Y}&=& -\bar Q'_{Li} Y_{Diq} q'_R \Phi_{q'} + h.c.\,,
 \ed
where $q'$ denotes the flavor of d-, s- and b-quark. Due to $\Phi_b$
being the next lightest scalar, in terms of mass eigenstates the
dominant effects for FCNCs at tree level in the $B$ processes are written
by
 \be
 {\cal L}_{\Delta B=1} &=& -\bar d_{Li} \left(V^{D}_{L}\right)_{ij}
 \left(Y_{D}\right)_{j3} b_R \phi^0_b \,. \label{eq:FCNC_L}
 \ed
 From the previous analysis, since the off-diagonal elements of
the flavor mixing matrix for the right-handed quark are small,
Eq.~(\ref{eq:FCNC_L}) only involves the flavor matrix matrix of
$V^{D}_L$.
Using Eq.~(\ref{eq:YQ}) and $V^{D}_{L} \approx
1+\Delta^D_L$, we see that
 \be
(V^D_L)_{ij}(Y_D)_{j3}\approx \lambda^D \delta_{ij} \delta_{j3} +
\delta_{ij}\epsilon^D_{j3} + \lambda^D (\Delta^{D}_L)_{ij}
\delta_{j3} +O(\epsilon^{D^2})\,.
 \ed
Furthermore, with the result of $\lambda^D (\Delta^D_L)_{j3} \approx
-\epsilon^D_{j3}$ shown in Eq.~(\ref{eq:delta}), we find that the
$\phi^0_b$-mediated FCNCs  not only are associated with the
parameter $\epsilon^D$, but also appear in $(\epsilon^D)^2$. As a
result, the contributions to the $B_q-\bar B_q$ mixing are
proportional to $(\epsilon^D)^4/m^2_{\phi^0_b}$. Clearly, by
choosing some suitable small value of $\epsilon^D$ and
$m_{\phi_b}\sim \rm TeV$,
they could be smaller than the current data. In other words, the
neutral Higgs mediated
$\Delta B=2$ processes will not provide a further constraint on the
parameters for the charged Higgs related effects.

 Now, we only pay attention to the charged Higgs related
effects. With the physical eigenstates of quarks, the charged Higgs
interactions to quarks can be found in Eq.~(\ref{eq:Yukawa}), given
by
 \be
 {\cal L}_{\CH}&=& - \bar u_{L} V_{CKM}
  \left[V^D_{L} Y_D\right] {\bf \Phi^{+}_{D}} d_{R} + \bar
u_{R} {\bf \Phi^{+}_{U}}  [V^{U}_{L} Y_U]^\dagger
V_{CKM} d_{L} + h.c\,, \label{eq:CH_nomass}
 \ed
where ${\bf \Phi^{+}_{F}}$ is a $3\times 3$ matrix and its
definition is similar to Eq.~(\ref{eq:Phi_F}). We note that ${\bf
\Phi^{+}_{F}}$ does not represent the physical charged Higgs
scalars.
Since there are six Higgs doublets in the model, basically we have 5
physical charged Higgs scalars and one charged Goldstone boson,
which  is usually  chosen to be
 \be
 G^{+}=\sum_{f=t,b,c,s,u,d}\frac{v_f}{v} \phi^{+}_{f}\,
 \ed
with $v=(\sum_{f} v^2_f)^{1/2}$. Therefore, to study the effects of
physical charged Higgses, in general, one needs to consider a
$6\times 6$ mass matrix for these charged scalar  fields. According
to our earlier analysis, the hierarchy of quark masses is
represented by the hierarchy of VEVs of the scalar fields. Due to $v_t\gg
v_{f\neq t}$, it should be a good approximation to take $v\approx
\sqrt{v^2_t+v^2_b +v^2_{c}}=\sqrt{2}(m^2_t + m^2_b +m^2_{c})^{1/2}
$, i.e., $\phi^{+}_{t}$ almost aligns to the Goldstone boson. Then,
the lightest charged Higgs will be the $\phi^{+}_{b}$. Moreover,
since $M_{b}< M_{c} \ll M_{s} \ll M_{d, u}$, the scalar mixing
effects associated with $\phi^{+}_{s, d, u}$ could be neglected due
to the suppression of their heavy masses. Based on the character of
the PH model, the interesting effects of the charged Higgses are in
fact only associated with $\phi^{+}_{t}$, $\phi^{+}_{b}$ and
$\phi^{+}_{c}$. Effectively, the charged Higgs  mass matrix is a
$3\times 3$ matrix, which is similar to that in the Weinberg
three-Higgs-doublet model \cite{Weinberg}. Interestingly, if we
further neglect the effect of $\phi^{+}_{c}$, the situation  returns
to the conventional two-Higgs-doublet model \cite{Lee}.
By using Eqs.~(\ref{eq:YQ}) and
(\ref{eq:delta}),
we obtain that diag($V^{Q}_{L} Y_Q)\approx (1,1,1)$.
 Moreover, from Eq.~(\ref{eq:CH_nomass}), we  find that
 the sizable effects due to the charged Higgs scalars
 are related to
 $\bar t_L b_R$ and $\bar t_R q_L$, where the vertex for the former is
given by $\sum^3_{k=1}(V_{CKM})_{3k} [V^D_{L} Y_D]_{k3}$ while the
latter  $\sum^3_{k=1}[V^U_L Y_U]^\dagger_{3k} (V_{CKM})_{k q}$. It
has no doubt that the coupling $\sum^3_{k=1}(V_{CKM})_{3k} [V^D_{L}
Y_D]_{k3}$ is dominated by k=3. However, it is more complicated for
the coupling $\sum^3_{k=1}[V^U_L Y_U]^\dagger_{3k} (V_{CKM})_{k q}$.
To see it,
we take $q=s$
with the sum
 \be
 \sum^3_{k=1}[V^U_L
Y_U]^\dagger_{3k} (V_{CKM})_{k s}&=&[V^U_L Y_U]^\dagger_{31}
(V_{CKM})_{u s}+[V^U_L Y_U]^\dagger_{32} (V_{CKM})_{c s}\non\\
&+& [V^U_L Y_U]^\dagger_{33} (V_{CKM})_{t s} \approx
\epsilon^{U\dagger}_{31} \lambda + \epsilon^{U\dagger}_{32}
+V_{ts}\,.
 \ed
In terms of
Eq.~(\ref{eq:delta}), the CKM matrix can be expressed by
$V_{CKM}=V^U_{L} V^{D^\dagger}_{L}\approx 1+\Delta^U_L -
\Delta^D_L$. Accordingly, we get $V_{ts}\approx
(\Delta^U_L)_{32} - (\Delta^U_L)_{32}=-\epsilon^U_{32}/\lambda_U +
\epsilon^D_{32}/\lambda_D$. Thus,
in the
phenomenological analysis, we can choose a suitable value of
$\lambda_{D(U)}$ so that $V_{ts} > \epsilon^{U\dagger}_{32}$.
The dominant effect for the vertex of $\bar t_L q_R$ could be
simplified to be $V_{ts}$, i.e., the 3-3 element of  $[V^{U}_{L} Y_U
]$ is the main contribution.

 In order to compare with the
conventional two-Higgs-doublet model, we rewrite
Eq.~(\ref{eq:CH_nomass}) in terms of quark masses  and
Eq.~(\ref{eq:vev_F}) as
 \be
 {\cal L}_{\CH}&=& -\sqrt{2}  \bar u_{L}V_{CKM}{\bf m_{D}  V^{-1}_{D}
 \Phi^{+}_{D}} d_{R} + \sqrt{2} \bar u_{R} {\bf \Phi^{+}_{U}m_{U}
  V^{-1}_{U} } V_{CKM} d_{L} +h.c. \label{eq:lang_ch}
 \ed
If we take ${\bf V^{-1}_{D(U)}}= {\bf \openone_{3\times3}}/v_{d(u)}$
and ${\bf \Phi^{+}_{D(U)}}= H^{+}_{d(u)} {\bf \openone_{3\times
3}}$, we can easily get the formulas for the charged-Higgs
interactions in the two-Higgs-doublet model to be
 \be
{\cal L}^{2\rm Higgs}_{\CH}&=& -\sqrt{2}  \bar u_{L}V_{CKM}{\bf
m_{D}} d_{R} \frac{H^+_d}{v_d}+ \sqrt{2} \bar u_{R} {\bf m_{U}}
   V_{CKM} d_{L} \frac{H^+_u}{v_u} +h.c. \label{eq:two-higgs}
 \ed
Furthermore, by using the relationships of
 \be
 G^{+}&=& \cos\beta H^{+}_{d} + \sin\beta H^{+}_{u}\,,\non\\
 H^{+}&=& -\sin\beta H^{+}_{d} + \cos\beta H^{+}_{u}\,
 \ed
with $\cos\beta=v_d/v$, $\sin\beta=v_u/v$ and $v=\sqrt{v^2_{d} +
v^2_u}$, we have
 \be
 {\cal L}^{2\rm Higgs}_{\CH}&=& (2\sqrt{2}G_F)^{1/2} \left( -\bar u_{L}V_{CKM}{\bf
m_{D}} d_{R} + \bar u_{R} {\bf m_{U}}
V_{CKM} d_{L} \right)G^{+} \non\\
&+& (2\sqrt{2}G_F)^{1/2} \left( \tan\beta \bar u_{L}V_{CKM}{\bf
m_{D}} d_{R} + \cot\beta \bar u_{R} {\bf m_{U}} V_{CKM} d_{L}
\right)H^{+}\,.
 \ed

\section{Phenomenologies in B decays}\label{sec:phe}

According to the discussions in Sec.~\ref{sec:PHM}, we  know that
there is an essential difference in the couplings of the charged
Higgs scalars and quarks between the conventional multi-Higgs and PH
models. For instance, if we turn off the CKM matrix elements, from
Eq.~(\ref{eq:CH_nomass}) we see clearly that the couplings in the
former are directly proportional to the masses of quarks
but those in the latter do not involve
new free parameters in the leading
contributions.
In addition, in the former case, there
are no intrinsic limits on the  charged Higgs masses, whereas in the
latter case, the masses have a preceding hierarchy stemmed from
Eq.~(\ref{eq:vev}). Consequently, we speculate that the lightest
charged Higgs scalar with the couplings of order  one in the PH
model might have interesting phenomenologies in rare decays
suppressed in the SM.
 From Eq.~(\ref{eq:CH_nomass}), one can easily find that the large novel
effects are associated with t and b quarks and the corresponding
charged Higgs scalars are mostly the first two lightest ones of
$\phi^{+}_{t}$ and $\phi^{+}_{b}$. Hence, in the following analysis,
we will concentrate on the rare $B$-meson processes
involving FCNCs due to the charged Higgs scalars.

\subsection{$B_{s,d}-\bar B_{s,d}$ mixings}\label{sec:mixing}

It is known that  all neutral pseudoscalar-antipseudoscalar
oscillations in the down type quark systems have been seen. In the
SM, since the oscillations are induced from box diagrams, they are
ideal  places to probe the new physics effects.
As
mentioned early, since $\phi^{+}_{s,d}$ are much heavier than
$\phi^{+}_{t,b}$, their contributions to the processes in the
$K$-system are small, whereas significant contributions   in the
B-system could be possible.

\begin{figure}[htbp]
\includegraphics*[width=5.0 in]{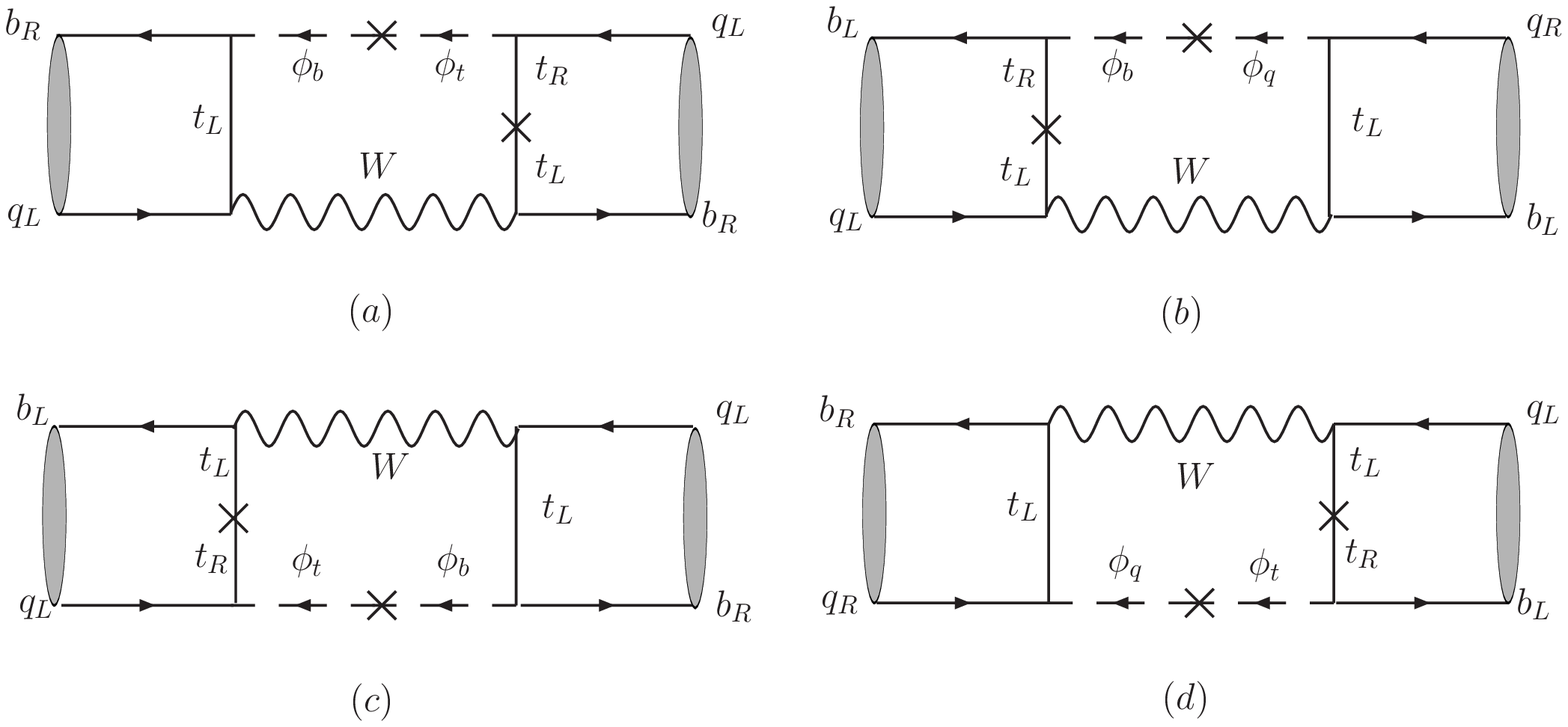}
\caption{Box diagrams for the $B_q-\bar B_q$ mixing induced by gauge
and charged Higgs bosons.}
 \label{fig:WH}
\end{figure}

\begin{figure}[htbp]
\includegraphics*[width=3.5
 in]{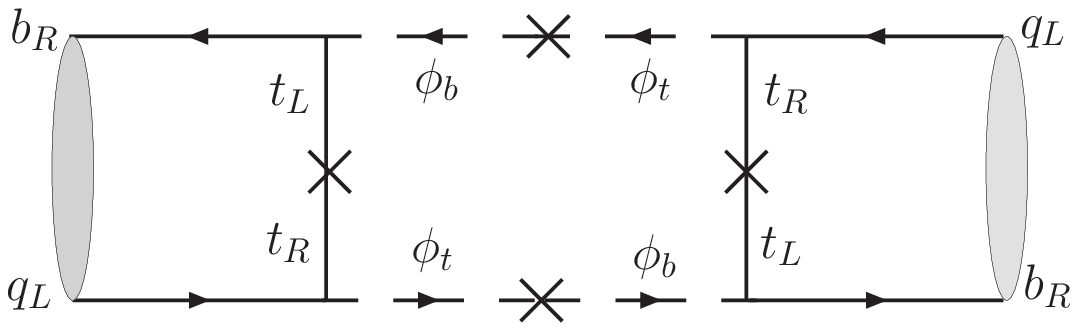}
\caption{Box diagrams for $B_q-\bar B_q$ mixing arisen from charged
scalar bosons.}
 \label{fig:HH}
\end{figure}

To calculate  $B_q-\bar B_q$ (q=d, s) mixings in the PH model, we
first consider the diagrams displayed in Fig.~\ref{fig:WH} due to
the gauge and charged Higgs bosons in the loop. The crossed diagrams
of internal bosons and fermions are included in the calculations but
not explicitly shown up in the figures. To see the mixing effects of
the charged Higgs scalars, we present the diagrams in terms of
unphysical states. However, we will formulate the results based on
the physical ones. Since Figs.~\ref{fig:WH}(b) and (d) involve the
heavy charged Higgs $\phi^{+}_{q}$, the contributions must be much
smaller than those by Figs.~\ref{fig:WH}(a) and (c) and therefore,
they can be ignored.
The effective four-fermion interactions for $\Delta B=2$ from
Figs.~\ref{fig:WH}(a) and (c) are given by
 \be
{\cal H}^{a}_{HW}&=& -\frac{G^2_F }{2\pi^2}\left(V_{tq} V^{*}_{tb}
\right)^2 m^2_{W} \left(\frac{m_b m_t}{m^2_{W}} C_{tb}  \right)
F\left(y_t, x_t\right) \bar b
\ga_{\mu} P_L q\, \bar b \ga^{\mu} P_L q \label{eq:HWbox}\,, \\
{\cal H}^{c}_{HW}&=&{\cal H}^{a}_{HW}\non
 \ed
 with
\be
F(a, b)&=& \int^{1}_{0} dx_1 \int^{x_1}_{0} dx_2\int^{x_2}_{0} dx_3
\frac{x_1-x_2+x_3}{[1-(1-a)x_1-(a-b)x_2]^2}\,,\non
 \ed
where $x_t=m^2_t/m^2_W$, $y_t=m^2_t/m^2_{H^+}$ and $C_{tb}$ denotes
the unknown mixing element between $\phi^{+}_t$ and $\phi^{+}_{b}$.
As discussed early, if we regard that the effective charged Higgs
scalars are $\phi^{+}_{t}$, $\phi^{+}_{b}$ and $\phi^{+}_{c}$, their
mixtures are similar to those in the Weinberg's three-Higgs-doublet
model. In general, $C_{tb}$ is a complex number. Here, for
simplicity, we have only shown the contributions of the lightest
physical charged Higgs denoted by $H^+$, referred as private charge
Higgs. Besides Fig.~\ref{fig:WH}, the diagram in Fig.~\ref{fig:HH}
also yields  an important contribution to the mixing.
 From Eq.~(\ref{eq:CH_nomass}),  we find
 \be
{\cal H}_{HH}&=& - \frac{G^2_{F}}{ \pi^2} \left( V_{td}
V^*_{tb}\right)^2 m^2_{W} \left( \frac{C_{tb}}{g^2}
\frac{m_{W}}{m_{\CH}} \frac{m_t}{m_{\CH}}\right)^2  G\left(
y_t\right)\bar b P_L q \bar b P_L q
 \ed
with
 \be
G(x)&=& -\frac{2}{(1-x)^2}-\frac{1+x}{(1-x)^3}\ln x\,\non.
 \ed
To examine the $B_q$ oscillating effect, we parametrize the matrix
elements as \cite{Masiero}
 \be
 \la B_q | (\bar q b)_{V-A} (\bar q b)_{V-A} | \bar B_q \ra &\approx&
\frac{4}{3} f^{2}_{B_q} \hat B_q m_{B_q}\,,\non \\
 \la B_q | (\bar q b)_{S+P} (\bar q b)_{S+P} | \bar B_q \ra &\approx&
-\frac{5}{6} f^{2}_{B_q} \hat B_q m_{B_q}\,,
 \ed
where $(\bar q b)_{V-A}=\bar q\ga_{\mu} (1-\ga_5)b$, $(\bar q
b)_{S+P}=\bar q (1+\ga_5) b$ and $f_{B_q}$ is the decay constant of
$B_q$. Accordingly, the $\bar B_q\to B_q$ matrix elements of ${\cal
H}_{ HW}$ and ${\cal H}_{ HH}$ are given by
 \be
 M^{qHW(H)}_{12}&=& \frac{G^2_F m^2_W}{12\pi^2} \left(V^{*}_{tq} V_{tb}
\right)^2 f^{2}_{B_q} \hat{B_q} m_{B_q} X_{HW(H)}\,,
\label{eq:MqH12}
%
 \ed
where
 \be
X_{HW}&=& -4\frac{m_b m_t}{m^2_W} C_{tb} F(y_t,x_t)\,, \non\\
X_{HH}&=& \frac{5}{2}\left(\frac{C_{tb}}{g^2} \frac{m_W
m_t}{m^2_{\CH}}\right)^2 G(y_t) \,,\label{eq:X}
 \ed
with $g$  the gauge coupling of $SU(2)_{L}$. We note that because
$X_{HW}$ has the suppression factor of $m_b/m_W F(y_t,x_t)$, it is
much smaller than $X_{HH}$. In the following analysis, we will
neglect the contribution of $X_{HW}$.

To study the influence of new physics on the time-dependent CPA, we
write the $\bar B_q\to B_q$ transition by combining results from the
SM and new physics as
 \be
M^{q}_{12}&=& A^{q SM}_{12} e^{-2i\beta_q } + A^{q NP}_{12}
e^{2i(\theta^{ NP}_{q}-\beta_q)} \label{eq:ms12}
 \ed
where $\beta_q\equiv arg(-V_{tq} V^*_{tb}/V_{cq}V^*_{cb})$ is the
weak CP phase of the SM, $\theta^{NP}_{q}$ corresponds to the new CP
phase in the PH model and $A^{qSM}_{12}$ is given by
 \be
A^{qSM}&=&\frac{G^2_F m^2_W}{12\pi^2} \left(V^{*}_{tq} V_{tb}
\right)^2 f^{2}_{B_q} \hat{B_q} m_{B_q} \eta_B S_{0}(x_t)
 \ed
with $\eta_B\approx 1$ and $S_{0}(x_t)\approx 0.784x_t^{0.76}$. Due
to $ \Delta \Gamma_q \ll\Delta m_q$ in the B-system \cite{PDG08},
the time-dependent CPA is found to be
 \be
- S_{J/\Psi M_q}&\simeq& {\rm
Im}\left(\sqrt{\frac{M^{q^*}_{12}}{M^{q}_{12}}}\right)
=\sin(2\beta_q - \phi^{\rm NP}_{q})\,, \non\\
\phi^{NP}_{q}&=& \arctan\left( \frac{r_q\sin2\theta^{\rm
NP}_{q}}{1-r_q \cos2\theta^{\rm NP}_{q}} \right) \label{eq:phiNPs}
 \ed
with $M_{d(s)}=K_S(\phi)$ and $r_q=A^{q\rm NP}_{12}/A^{q\rm
SM}_{12}$. From Eqs.~(\ref{eq:MqH12}) and (\ref{eq:X}), one gets
that $\theta^{NP}_{q}\equiv\theta_{\CH}=arg(C_{tb})$ and
 \be
 r_{q}&\equiv& r_H= \frac{|X_{HH}|}{\eta_B S_0(x_t)}\,,
 \ed
which is independent of $q$ in the PH model.
 From Eq.~(\ref{eq:phiNPs}), it is readily seen that
the magnitude of $\phi^{NP}_{s}$ is controlled by $r_H$.

\subsection{$b\to q \ga$ decays}

\begin{figure}[htbp]
\includegraphics*[width=5.0
 in]{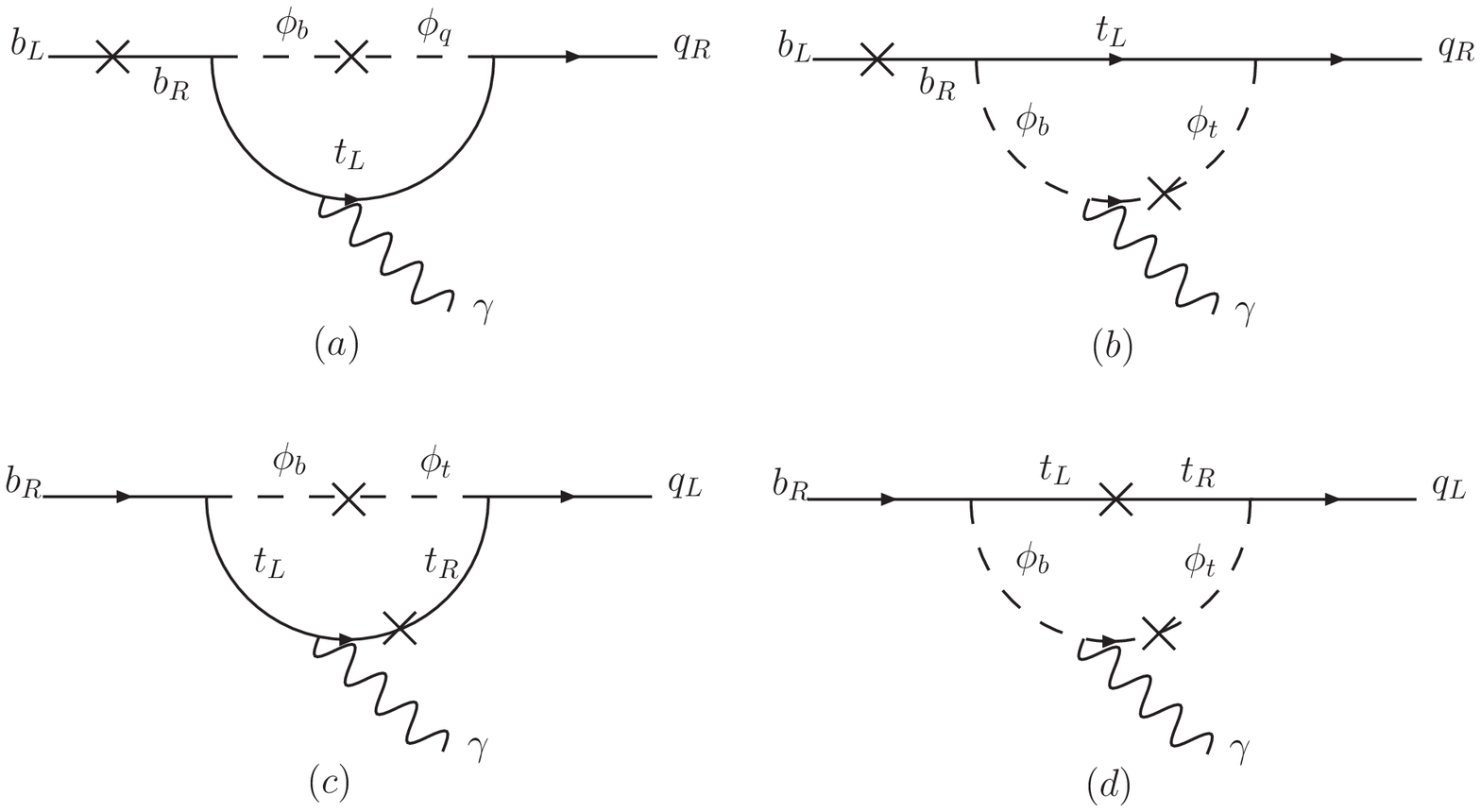}
\caption{Penguin diagrams for $b\to q \ga$ decays by charged Higgs
scalars in the PH model.}
 \label{fig:penguin}
\end{figure}

It is known that $b\to q\ga$ decays provide strong constraints on
the penguin contributions from new physics. In this subsection, we
examine these decays in the PH model. As an illustration, we present
the possible dominant effects in Fig.~\ref{fig:penguin}.
 From the figure, we see clearly that Figs.~\ref{fig:penguin}(a) [(c)] and
(b) [(d)] involve chirality flip of $b$ [$t$]  and the mixing of
$\phi^+_b$ and $\phi^+_q$ [$\phi^+_t$]. Due to $m_b\ll m_t$ and the
mixing effect of $\phi^+_{b}$ and $\phi^+_{q}$ ($\propto
\ga_{qb}/M^2_{\phi_q}$) being much smaller than that of $\phi_b$ and
$\phi_t$ ($ \propto \ga_{tb}/m^2_W$), the contributions of
Figs.~\ref{fig:penguin}(a) and (b) are much smaller than those of
Figs.~\ref{fig:penguin}(c) and (d). Therefore, to study the leading
effects, the results of Figs.~\ref{fig:penguin}(a) and (b) can be
neglected. Furthermore, if we replace photons in
Fig.~\ref{fig:penguin} with gluons, gluonic penguins can be also
generated by the charged Higgs scalars in the PH model.

 From Figs.~\ref{fig:penguin}(c) and (d), we conclude
 that the
effective operators
from the  charged scalars have the same structures as those in the
SM. In order to include the SM contributions, we write the effective
Hamiltonian for $b\to q\ga$ as \cite{BBL}
 \be
 {\cal H}(b\to q\ga)&=&-\frac{G_F}{\sqrt{2}} V^{*}_{tq} V_{tb}
 \left[ \sum^{6}_{i=1} C_{i}(\mu) O_{i}(\mu) + C_{7\ga}(\mu) O_{7\ga}(\mu)
 + C_{8G}(\mu) O_{8G}(\mu)\right]\,,
 \ed
where  $O_{i}(\mu)$ are the  effective operators at $\mu$ scale and
$C_{i}(\mu)$ are the corresponding Wilson coefficients. Because the
dominant effects of the SM are from the terms with $C_{2}$,
$C_{7\ga}$ and $C_{8G}$, we only show the associated operators of
 \be
 O_{2}&=&(\bar q c)_{V-A} (\bar c b)_{V-A}\,,\non\\
 O_{7\ga}&=& \frac{e}{8\pi^2} m_b \bar q \sigma^{\mu \nu}(1+\ga_5) b
 F_{\mu\nu}\,,\\
O_{8G}&=& \frac{g_s}{8\pi^2} m_b \bar q_{\alpha} \sigma^{\mu
\nu}(1+\ga_5) T^{a}_{\alpha \beta} b_{\beta}
 G^{a}_{\mu\nu}\,,
 \ed
 respectively,
where $(\bar f f')_{V-A}=\bar f \ga_{\mu} (1-\ga_5) f'$, $e$ is the
electric charge, $g_s$ is the strong coupling constant, $\alpha$ and
$\beta$ denote the color indices, $T^{a}_{\alpha \beta}$ with
a=1,\ldots,8 are the generators of the $SU(3)_C$ gauge symmetry and
$F_{\mu\nu}$  ($G^{a}_{\mu\nu}$) is the electromagnetic (gluonic)
field strength. The effective Wilson coefficients by combining the
contributions of the W-boson and lightest charged Higgs are given by
 \be
 C_{7\ga,\, 8G}&=& C^{W}_{7\ga,\, 8G} + C^{H^+}_{7\ga,\, 8G}
 \ed
with
 \be
C^{\CH}_{7\ga}&=& \frac{\rm v^2}{4m^2_{\CH}} \frac{m_t}{m_{b}}
C_{tb} \left( Q_t I_c\left( y_t\right)
+I_d\left(y_t\right)\right)\,,\non\\
 C^{\CH}_{8G}&=& \frac{\rm v^2}{4m^2_{\CH}} \frac{m_t}{m_{b}} C_{tb} I_d\left(
 y_t \right)\,, \label{eq:c7ga}
 \ed
where $C^{W}_{7\ga(8G)}$ denotes the SM result, $Q_t$ is the
electric charge of the top quark and the loop integrals $I_{c}$ and
$I_{d}$  come from Figs.~\ref{fig:penguin}(c) and (d), given by
 \be
 I_{c}(x)&=& -\frac{3-x}{2(1-x)^2}-\frac{1}{(1-x)^3} \ln x\,,
 \non\\
 I_{d}(x)&=& \frac{1+x}{2(1-x)^2} + \frac{x}{(1-x)^3}\ln x\,,
 \ed
respectively.

\subsection{$B_q\to \ell^{+} \ell^{-}$ and $B\to (P,\, V) \ell^{+} \ell^{-}$ decays}

\begin{figure}[htbp]
\includegraphics*[width=5.0
 in]{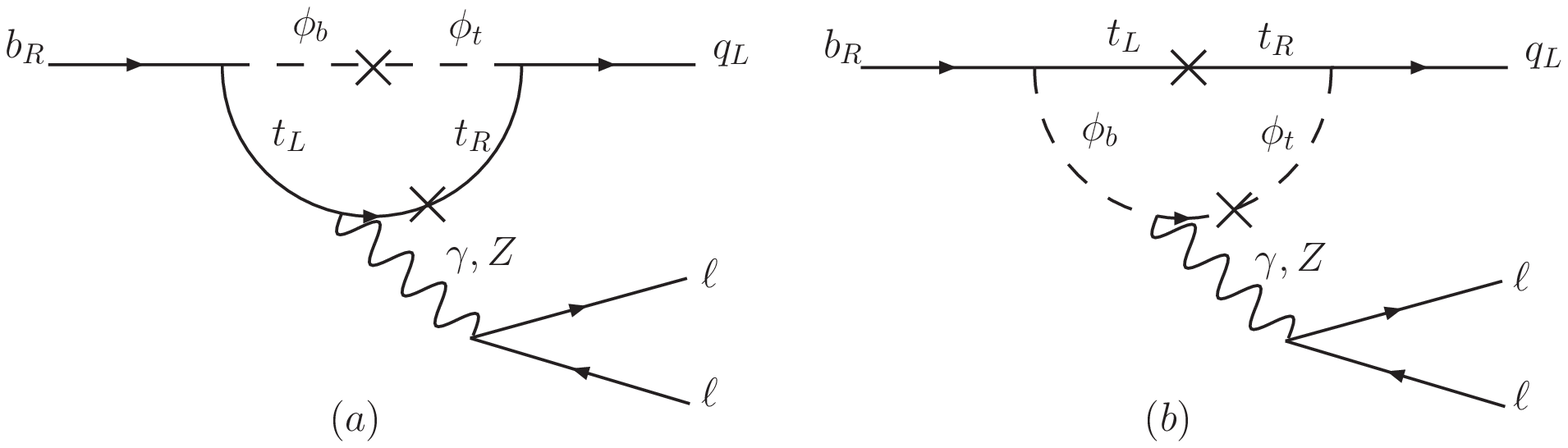}
\caption{Penguin diagrams for $b\to q \bar \ell \ell$ decays
generated by the charged scalars in the PH model.}
 \label{fig:bqll}
\end{figure}

In this subsection, we discuss the leptonic $B_q\to \ell^{+}
\ell^{-}$ and semileptonic $B\to (P,\, V)\ell^{+} \ell^{-}$ decays.
The effective Hamiltonian for $b\rightarrow q \ell^{+}\ell^{-}$ in
the SM is given by~\cite{BBL,CG_NPB636,CG_PRD66}
 \be
{\cal H}(b\to q \ell^{+} \ell^{-} )= -\frac{G_{F}\alpha
}{\sqrt{2}\pi } \lambda^{q}_t \left[ H_{1\mu }L^{\mu }+H_{2\mu
}L^{5\mu }\right] \label{eq:heff}
 \ed
with
 \be
H_{1\mu } &=&C^{\rm eff}_{9}(\mu )\bar{q}\gamma _{\mu }(\mu )P_{L}b\
-\frac{2m_{b}}{ k^{2}}C^{W}_{7\ga}(\mu )\bar{q}i\sigma _{\mu \nu
}k^{\nu }P_{R}b \,,
\non\\
H_{2\mu } &=&C_{10}\bar{q}\gamma _{\mu }P_{L}b \,,
\non\\
L^{\mu } &=&\bar{\ell}\gamma ^{\mu }\ell\,, \non\\
L^{5\mu } &=&\bar{\ell}\gamma ^{\mu }\gamma _{5}\ell\,,
 \ed
where $\lambda^{q}_{t}=V^{*}_{tq} V_{tb}$, $k^2$ is the invariant
mass of the lepton-pair and $C^{\rm eff}_{9}(\mu )$, $C_{10}$ and
$C^{W}_{7\ga}(\mu )$ are the Wilson coefficients (WCs) with their
expressions for next leading order corrections
in Ref. \cite{BBL}.  Since the operator associated with $C_{10}$ is
not renormalized under QCD, it is the only one with the $\mu $ scale
free. In addition, by considering the effects from the one-loop
matrix elements of  $O_{1}=\bar{s}_{\alpha }\gamma ^{\mu
}P_{L}b_{\beta }\ \bar{c}_{\beta }\gamma _{\mu }P_{L}c_{\alpha }$
and $ O_{2}=\bar{s}\gamma ^{\mu }P_{L}b\ \bar{c}\gamma _{\mu
}P_{L}c$, the resultant effective WC of $C_{9}$ is \cite{BBL}
 \be
C_{9}^{\rm eff}&=&C_{9}\left( \mu \right) +\left( 3C_{1}\left( \mu
\right)
+C_{2}\left( \mu \right) \right)  h( x,s) \,, \non\\
h(z,s)&=&-\frac{8}{9}\ln\frac{m_b}{\mu}-\frac{8}{9}\ln z
+\frac{8}{27} +\frac{4}{9}x  -\frac{2}{9}(2+x)|1-x|^{1/2} \nonumber
\\
&\times& \left\{
  \begin{array}{c}
\ln \left|\frac{\sqrt{1-x}+1}{\sqrt{1-x}-1} \right|-i\, \pi, \
{\rm for\ x\equiv 4z^2/s<1 }\, , \\
2\, \arctan\frac{1}{\sqrt{x-1}},\   {\rm for\ x\equiv 4z^2/s>1 }  \\
  \end{array}
\right.\label{effc9}
 \ed
with $z=m_c/m_b$ and $s=k^2/m^2_B$. Similar to the SM, electroweak
penguin diagrams in Fig.~\ref{fig:bqll} mediated by the private
charged Higgs scalars can also contribute to $b\to q \ell^{+}
\ell^{-}$. Therefore, in terms of Eq.~(\ref{eq:CH_nomass}) and the
mixture of $\phi^{+}_{t}$ and $\phi^{+}_{b}$, the results of Z- and
$\ga$-penguin are formulated to be
 \be
{\cal H}^{Z}_{a+b}&=& \frac{G_F \alpha}{\sqrt{2}\pi} \lambda^{q}_t
\left\{ -X^{Z}_{1} \bar q \ga_{\mu} P_L b \left[C^{\ell}_{V}
\bar\ell
\ga^{\mu} \ell -C^{\ell}_{A} \bar\ell \ga^{\mu} \ga_{5} \ell \right] \right. \non \\
&+&\left. X^{Z}_{2} \bar q P_R b \left[ C^{\ell}_{V} \bar\ell \not
P_{b} \ell -
C^{\ell}_{A} \bar \ell \not P_b \ga_{5} \ell \right]\right\}\,, \non \\
{\cal H}^{\ga}_{a+b} &=& \frac{G_F \alpha}{\sqrt{2}\pi}
\lambda^{q}_{t} \left\{ Y^{\ga} \bar q\ga_{\mu} P_L b \bar\ell
\ga^{\mu} \ell + C^{\CH}_{7\ga}  \frac{2m_b}{k^2} \bar q
i\sigma_{\alpha\beta} k^{\beta} P_R b  \bar\ell \ga^{\alpha}  \ell
\right\}\,, \label{eq:Zga}
 \ed
 respectively,
with
 \be
X^{Z}_{1} &=&\frac{C_{tb} m_b}{8\pi \alpha}
\frac{m_t}{m^2_{H}}\left[
\left(C^t_V + C^t_A\right) K_1(y_t) + \frac{1}{2}\left( C^t_V - C^t_A\right) I_{c}(y_t) \right]\,, \non \\
 X^{Z}_{2} &=&\frac{C_{tb}}{4\pi \alpha}
\frac{m_t}{m^2_{H}}\left( C^t_V I_c(y_t) +\cos2\theta_W I_d(y_t)
\right)\,,\non\\
Y^{\ga} &=& \frac{C_{tb}}{2}\frac{{\rm v}^2}{m^2_{\CH}} \frac{m_b
m_t}{k^2} (1-Q_t) I_d(y_t)\,,\non\\
C^{f}_{V} &=& T^{3}_{f}-2\sin^2\theta_W Q_{f},\ \ \
C^{f}_{A}=T^{3}_{f} \label{eq:XXY}
 \ed
and
 \be
 K_{1}(x)&=& -\frac{1-3x}{4(1-x)^2}-\frac{1-x}{2(1-x)^3}\ln x\,,
 \ed
where $T^3_{f}$  is the third component of weak isospin and $Q_{f}$
is the electric charge of  $f$.

With the effective interactions  in Eqs.~(\ref{eq:heff}) and
(\ref{eq:Zga}) for $b\to q \ell^{+} \ell^{-}$, the BR for the
two-body decay $B_q\to \ell^{+} \ell^{-}$ is straightforwardly given
by
 \be
{\cal B}(B_q\to \ell^+ \ell^-)&=& {\cal B}^{SM}(B_q\to \ell^+
\ell^-) \left| 1+\frac{C^{\ell}_{A} X^{Z}_{1}}{C_{10}} +
+\frac{C^{\ell}_{A} X^{Z}_{2}}{C_{10}}\frac{m^2_{B_q}}{m_b}
\right|^2 \label{eq:BrBqll}
 \ed
with
 \be
  {\cal B}^{SM}(B_q\to \ell^+ \ell^-) &=& \tau_{B_q}\frac{G^2_F
\alpha^2}{16\pi^3} |\lambda^{q}_{t}|^2 m_{B_q}f^2_{B_q}m^2_{\ell}
\left( 1-\frac{4m^2_{\ell}}{m^2_{B_q}}\right)^{1/2}
|C_{10}|^2\,.\non
 \ed
Since the BR is proportional to the lepton mass, obviously, the
related decays are chiral suppressed. In addition, we see that only
the $\CH$ mediated Z-penguin has the contribution to the decays. In order to study
$B\to (P,\, V)\ell^{+} \ell^{-}$, we have to
know the information on the transition elements of $B\rightarrow
\left( P,\ V\right) $ with various transition currents. As usual, we
parametrize the relevant form factors as follows:
 \be
\langle \bar P(p_{P})| V_{\mu }| \bar{B} (p_{B})\rangle &=&
f_{+}(k^2)\Big\{P_{\mu}-\frac{P\cdot k }{k^2}k_{\mu} \Big\}
+\frac{P\cdot k}{k^2}f_{0}(k^2)\,k_{\mu},  \non \\
\langle \bar P(p_{P} )| T_{\mu \nu}k^{\nu}| \bar{B} (p_{B})\rangle
&=& \frac{f_{T}(k^2)}{ m_{B}+m_{P}}\Big\{P\cdot k\,
k_{\mu}-k^{2}P_{\mu}\Big\},
\non  \\
\langle \bar V(p_{V},\epsilon )| V_{\mu }| \bar{B} (p_{B})\rangle
&=&i\frac{V(k^{2})}{m_{B}+m_{V}}\varepsilon _{\mu
\alpha \beta \rho }\epsilon ^{*\alpha }P^{\beta }k^{\rho },  \non \\
\langle \bar V(p_{V},\epsilon )| A_{\mu }| \bar{B} (p_{B})\rangle
&=&2m_{V}A_{0}(k^{2})\frac{\epsilon ^{*}\cdot k}{ k^{2}}q_{\mu }+(
m_{B}+m_{V}) A_{1}(k^{2})\Big( \epsilon
_{\mu }^{*}-\frac{\epsilon ^{*}\cdot k}{k^{2}}k_{\mu }\Big)  \non \\
&&-A_{2}(k^{2})\frac{\epsilon ^{*}\cdot k}{m_{B}+m_{V}}\Big( P_{\mu
}-\frac{P\cdot k}{k^{2}}k_{\mu }\Big) ,  \non \\
\langle \bar V(p_{V},\epsilon )| T_{\mu \nu }k^{\nu }| \bar{B}
(p_{B})\rangle &=&-iT_{1}(k^{2})\varepsilon _{\mu \alpha \beta \rho
}\epsilon ^{*\alpha }P^{\beta }k^{\rho },  \non \\
\langle \bar V(p_{V},\epsilon )| T_{\mu \nu }^{5}k^{\nu }|
\bar{B}(p_{B})\rangle &=& T_{2}(k^{2})\Big( \epsilon _{\mu
}^{*}P\cdot k-\epsilon ^{*}\cdot k P_{\mu }\Big)
+T_{3}(k^{2})\epsilon ^{*}\cdot k \Big( k_{\mu }-\frac{k^{2}}{P\cdot
k}P_{\mu }\Big)\, , \label{ffv}
 \ed
where $(V_{\mu },A_{\mu },T_{\mu \nu },T_{\mu \nu
}^{5})=\bar{q}(\gamma _{\mu }, \gamma_{\mu }\gamma _{5}, i\sigma
_{\mu \nu }, i\sigma _{\mu \nu }\gamma _{5})b$, $m_{B,P,V}$ are the
masses of $B$, pseudoscalar and vector mesons, $P=p_{B}+p_{P(V)}$,
respectively, $k=p_{B}-p_{P(V)}$ and $P \cdot
k=m^{2}_{B}-m^{2}_{P(V)}$. By equation of motion, we can have the
transition form factors for scalar and pseudoscalar currents as
  \be
\langle \bar P| \bar{q} \, b| \bar{B} \rangle & \approx &
\frac{P\cdot
k}{m_b}\, f_0(k^2),\non \\
\langle V| \bar{q} \gamma_5 b| \bar{B} \rangle & \approx & -
\frac{2m_{V}}{m_b } \epsilon^*\cdot q\, A_0(k^2)\,.
    \label{sff}
  \ed
Here,  the light quark mass has been neglected. According to the
definitions of the form factors, the transition amplitudes for
$B\rightarrow (P,\, V) \ell^{+}\ell^{-}$ can be written as
 \be
{\cal M}_{P}&=&-\frac{G_{F}\alpha \lambda^{q}_{t}}{\sqrt{2}\pi }
\left[ m_{97} \bar{\ell} \not{p}_K \ell + m_{10} \bar{\ell}
\not{p}_K \gamma_5 \ell  \right] \label{eq:tampk}
 \ed
 with
 \be
m_{97}&=& \left(C^{\rm eff}_9+ C^{\ell}_{V} X^{Z}_{1}-Y^{\ga}\right)
f_+ - \frac{m_B}{2} C^{\ell}_{V} C^{Z}_{\CH} f_{0}
+\frac{2m_b}{m_B+m_P}C_{7\ga}
f_T \,,\non \\
m_{10}&=& \left(C_{10}-C^{\ell}_{A} X^{Z}_{1}\right) f_+ +
\frac{m_B}{2} C^{\ell}_{A} C^{Z}_{\CH} f_{0}\,,
  \label{eq:TMK}
 \ed
and
\begin{equation}
{\cal M}_{V}=-\frac{G_{F}\alpha \lambda^{q}_{t}}{\sqrt{2}%
\pi }\left[ {\cal M}_{1\mu }\bar{\ell}\gamma^{\mu} \ell+{\cal
M}_{2\mu } \bar{\ell}\gamma^{\mu}\gamma_5 \ell \right]
\label{eq:ampk*}
\end{equation}
where
\begin{eqnarray}
{\cal M}_{1\mu } &=& i\tm^{1}_{97}\varepsilon _{\mu \nu \alpha \beta
}\epsilon ^{*\nu }p^{\alpha }_{V} k^{\beta }-\tm^2_{97} \epsilon
_{\mu }^{*}+\tm^{3}_{97}\epsilon ^{*}\cdot k p_{V\mu },
\non \\
{\cal M}_{2\mu } &=& i\tm^{1}_{10}\varepsilon _{\mu \nu \alpha \beta
}\epsilon ^{*\nu }p^{\alpha }_{V} k^{\beta }-\tm^2_{10} \epsilon
_{\mu }^{*}+\tm^{3}_{10}\epsilon ^{*}\cdot k p_{V\mu },
 \label{eq:ampk1}
\end{eqnarray}
with
 \be
\tm^{1}_{97}&=& \frac{V}{m_B+m_{V}} \left(C^{\rm eff}_9 + C^{\ell}_V
X^{Z}_{1}-Y^{\ga}\right)
+\frac{2m_b}{k^2}C_{7\ga} T_1 \, ,\non\\
\tm^{2}_{97}&=& \frac{1}{2}(m_B+m_{V}) \left(C^{\rm eff}_9 +
C^{\ell}_V X^{Z}_{1}-Y^{\ga}\right) A_1
+\frac{1}{2} \frac{2m_b}{k^2}P\cdot k C_{7\ga} T_2 \, ,\non\\
\tm^{3}_{97}&=& \frac{A_2}{m_B+m_{V}} \left(C^{\rm eff}_9 +
C^{\ell}_V X^{Z}_{1}-Y^{\ga}\right) + \frac{m_{V}}{m_B}
A_{0}C^{\ell}_{V} C^{Z}_{\CH}
+\frac{2m_b}{k^2}C_{7\ga} \left(T_2+\frac{k^2}{P\cdot k}T_3\right) \, ,\non\\
\tm^{1}_{10}&=& \frac{V}{m_B+m_{V}} \left(C_{10}-C^{\ell}_{A} X^{Z}_{1}\right)\, ,\non\\
\tm^{2}_{10}&=& \frac{1}{2}(m_B+m_{V})A_1 \left(C_{10}-C^{\ell}_{A} X^{Z}_{1}\right) \, ,\non\\
\tm^{3}_{10}&=& \frac{A_2}{m_B+m_{V}} \left(C_{10}-C^{\ell}_{A}
X^{Z}_{1}\right) - \frac{m_{V}}{m_B} A_{0} C^{\ell}_{A} C^{Z}_{\CH}
\, .
 \label{TMKs}
 \ed
Here, we only pay attention to the light leptons with the explicit
effects of $m_{\ell}$ ignored.

To get the decay rate distribution in terms of the dilepton
invariant mass $k^2$ and the lepton polar angle $\theta$, we use the
$k^2$ rest frame in which $p_{\ell}=E_{\ell}(1, \sin\theta, 0,
\cos\theta)$, $p_{H}=(E_{H},0,0,|\vec{p}_{H}|\cos\theta)$ with
$E_{\ell}=\sqrt{k^2}/2$,
$E_{H}=(m^{2}_{B}-k^2-m^2_{H})/(2\sqrt{k^2})$ and
$|\vec{p}_{H}|=\sqrt{E^2_{H}-m^{2}_{H}}$. By squaring the transition
amplitude in Eq.~(\ref{eq:tampk}) and including the three-body phase
space factor, the differential decay rate as a function of $k^2$ and
$\theta$ for $B\to P \ell^+ \ell^-$ is given by
 \begin{eqnarray}
\frac{d\Gamma _{P} }{dk^2 d\cos\theta}&=&\frac{G_{F}^{2}\alpha^{2}|
\lambda^{q} _{t}| ^{2}}{ 2^{8}m^{2}_B \pi ^{5}}\tilde{p}_{P}
|\vpp|^2 \left(k^2-4E_{\ell}^2 \cos^2\theta \right)\left(
|m_{97}|^2+|m_{10}|^2\right) . \label{eq:difk}
 \end{eqnarray}
For $B\rightarrow V \ell^{+} \ell^{-}$, by summing up the
polarizations of $V$ with the identity $\sum \e^{*}_{V\mu}(p)
\e_{V\nu}(p) = (-g_{\mu\nu}+p_{\mu}p_{\nu}/p^2)$, from Eq.
(\ref{eq:ampk*}) the differential decay rate is found to be
 \begin{eqnarray}
\frac{d\Gamma_{V} }{ dk^{2}d\cos \theta} &=&\frac{
G_{F}^{2}\alpha^{2}|\lambda _{t}| ^{2} }{2^{8}m_{B}^{2} \pi ^{5}}
\tilde{p}_{V}   \left\{ k^2 |\vpv|^2
\left(k^2+4E_{\ell}^2\cos^2\theta \right)
\left(|\tm^1_{97}|^2+|\tm^1_{10}|^2 \right) \right. \non
\\
&+&   \frac{ |\vpv|^2}{m^2_{V}}  \left( k^2-4 E_{\ell}^2
\cos^2\theta \right) \left(|\tm^2_{97}|^2+|\tm^2_{10}|^2 \right) +
2k^2 \left(|\tm^2_{97}|^2 + |\tm^2_{10}|^2
\right) \non \\
&+&  \frac{k^2}{m^2_{V}} |\vpv|^4 \left( k^2 -4E_{\ell}^2
\cos^2\theta \right) \left( |\tm^3_{97}|^2 + |\tm^3_{10}|^2
\right) \non \\
&-&  2\frac{k\cdot p_{V}}{m^2_{V}} |\vpv|^2 \left( k^2 -4E_{\ell}^2
\cos^2\theta \right) \left( Re(\tm^2_{97} \tm^{3*}_{97}) +
Re(\tm^2_{10}\tm^{3*}_{10})\right) \non \\
&-& \left.8 |\vpv|E_{\ell} k^2 \left[Re(\tm^1_{97}\tm^{2*}_{10})+
Re( \tm^2_{97}\tm^{1*}_{10}) \right] \cos\theta
  \right\}\label{eq:difks}.
\end{eqnarray}
Here, $\tilde{p}_{H}$ ($H=p$ or $V$) represents the spatial momentum
of the $H$ meson in the $B$-meson rest frame, given by
$\tilde{p}_{H}=\sqrt{ E^{\prime 2}_{H}- m^{2}_{H} }$ with $E^{\prime
}_{H}=(m^{2}_{B}+m^{2}_{H}-k^{2})/(2m_{B})$. The forward-backward
asymmetry (FBA) is defined by
 \be
A_{FB}&=& \frac{\int^{1}_{-1} \omega(\theta)d\cos\theta
d\Gamma/dk^2/d\cos\theta}{\int^{1}_{-1} d\cos\theta
d\Gamma/dk^2/d\cos\theta} \label{eq:FBA}
 \ed
with $\omega(\theta)=\cos\theta/|\cos\theta|$. Since
Eq.~(\ref{eq:difk}) has no linear term in $\cos\theta$, the FBA for
$B\to P \ell^{+} \ell^{-}$ vanishes. Hence, only $B\to V \ell^{+}
\ell^{-}$ has a nonvanished FBA, given by
 \be
A_{FB}^{V}(k^2) &=&- \frac{1}{d\Gamma/dk^2 }\frac{
G_{F}^{2}\alpha^{2}|\lambda _{t}| ^{2} }{2^{8}m_{B}^{2} \pi
^{5}} \tilde{p}_{V} \non\\
&\times& \left[8 |\vpv|E_{\ell} k^2
\left(Re(\tm^1_{97}\tm^{2*}_{10})+ Re( \tm^2_{97}\tm^{1*}_{10})
\right)\right]\,.
 \ed

\section{Numerical results and discussions}\label{sec:num}

Since the contributions to the processes in the $B_q$ mixing, $B\to
X_s \ga$, $B_q\to \ell^{+} \ell^{-}$ and $B\to (P,\, V) \ell^{+}
\ell^{-}$  by the charged Higgs scalars have strong correlations,
the new free parameters are only $m_{\CH}$ and $C_{tb}$. On the
other hand, we can  find constraints among these decays due to
experimental data. To comprehend the influence of the new charged
Higgs on the rare decays, we in turn investigate the above
processes. As an illustration, we only focus on the  processes with
$\ell=\mu$.

\begin{figure}[bpth]
\includegraphics*[width=4 in]{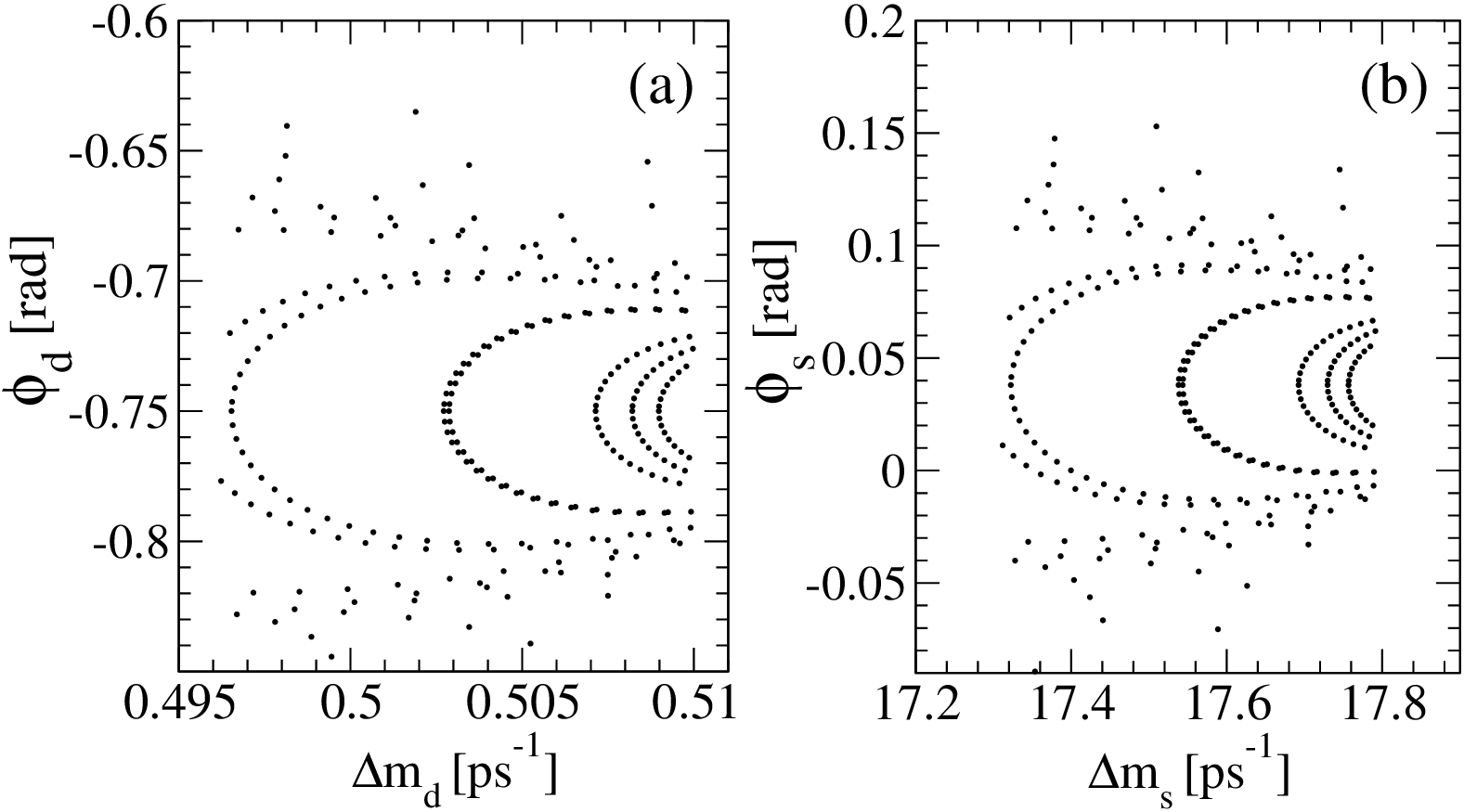}
\caption{(a)[(b)] $\phi_{d[s]}=2\beta_{d[s]}-\phi^{\CH}_{d[s]}$
versus $\Delta m_{d[s]}$.}
 \label{fig:phi-mq}
\end{figure}

For the $B_q$ mixing, besides the mass difference of two physical
B-meson states described by  $\Delta m_q=2|M^{q}_{12}|$, the
time-dependent CPA  in Eq.~(\ref{eq:phiNPs}) is also an important
physical quantity to display the new physics. To do the numerical
analysis, we take $f_{B_d}\sqrt{\hat B_d}=0.184$ GeV,
$f_{B_s}\sqrt{\hat B_s}=0.221$ GeV, $V_{ts}=-0.04e^{i\beta_s}$ with
$\beta_s=0.019$ and $V_{td}=8.2\times 10^{-3} e^{-i\beta_d}$ with
$\beta_s=-0.375$ \cite{HFAG}, in which the leading SM results are
$\Delta m^{SM}_{d}=0.52$ ps$^{-1}$ and $\Delta m^{SM}_{s}=18.25$
ps$^{-1}$. Accordingly, we present the influence of the private
charged Higgs in Fig.~\ref{fig:phi-mq}, where
$\phi_{q}\equiv2\beta_q - \phi^{\CH}_{q}$, $m_{\CH}$ is set to be
150 GeV $\le m_{\CH}\le$ 1 TeV and $C_{tb}$ and $\theta_{\CH}$ have
been chosen to satisfy $0.49\le \Delta m_d \le 0.51$ ps$^{-1}$ and
$17.31\le \Delta m_s \le 19.03$ ps$^{-1}$ \cite{CDF,D0}. We note
that although $\Delta m_d$ has a very high precise measurement with
$0.507 \pm 0.055$ ps$^{-1}$ \cite{PDG08}, since the error from the
nonperturbative QCD is large, for theoretical estimations we take a
conservative bound. From the figure, we see clearly that if we only
consider the constraints of $\Delta m_{d,s}$, the CP phases
extracted from time-dependent CPAs of $B\to J/\Psi (K_S,\, \phi)$
have significant deviations from those in the SM.

\begin{figure}[bpth]
\includegraphics*[width=4 in]{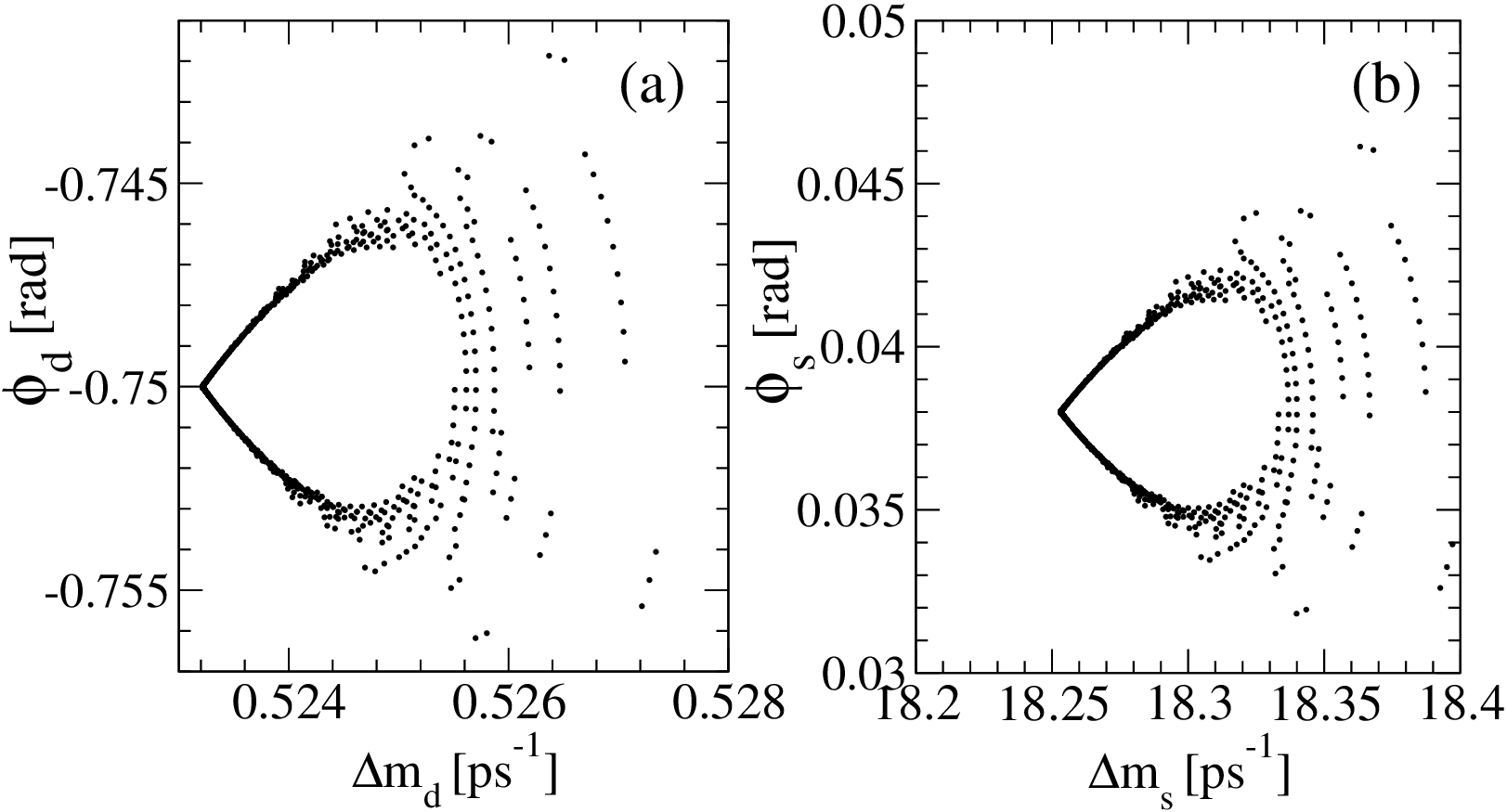}
\caption{(a)[(b)] $\phi_{d[s]}=2\beta_{d[s]}-\phi^{\CH}_{d[s]}$
versus $\Delta m_{d[s]}$ while the limits of ${\cal B}(B\to
X_s\ga)=(3.52\pm 0.25)\times 10^{-4}$ and $A_{CP}(b\to
s\ga)=0.004\pm 0.037$ with 1$\sigma$ errors are considered.}
 \label{fig:phi-mq-bsga}
\end{figure}

\begin{figure}[bpth]
\includegraphics*[width=4 in]{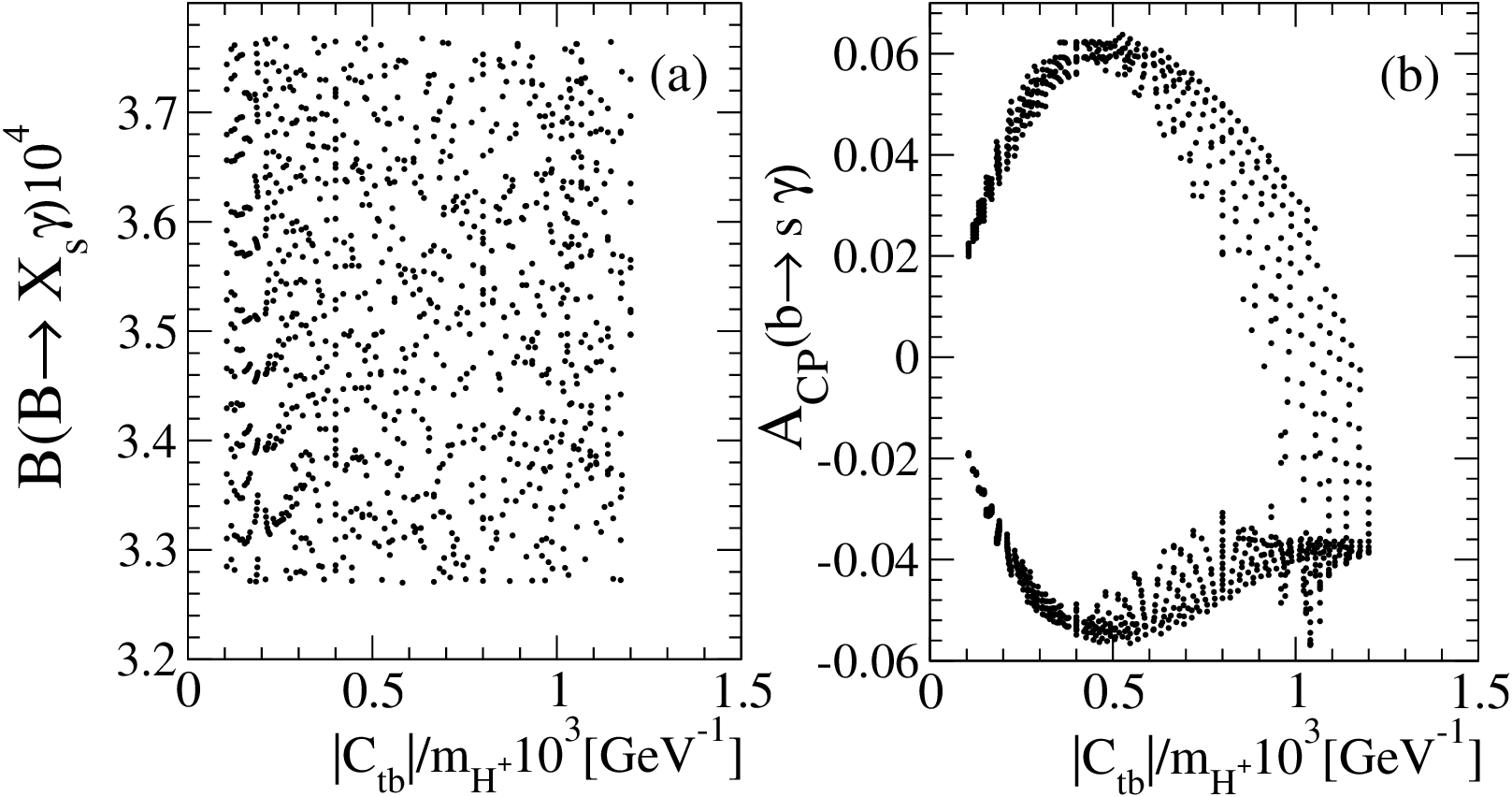}
\caption{(a)[(b)] Correlation between ${\cal B}(B\to X_s\ga)$ (in
units of $10^{-4}$) [$A_{CP}(b\to s\ga)$] and parameter
$|C_{tb}|/m_{\CH}$ (in units of $10^{-3}$).}
 \label{fig:ctb-mh}
\end{figure}

It is known that the BR for $B\to X_s \ga$ not only has been
measured well to be $(3.52\pm 0.23\pm 0.09)\times 10^{-4}$
\cite{HFAG} but also is consistent with the SM prediction of $(3.29
\pm 0.33)\times 10^{-4}$ \cite{bsgaSM}. Hence, $B\to X_s\ga$ could
give a strict constraint on the parameters of new physics. To simply
get the bound, we adopt the BR for $B\to X_s\ga$ to be \cite{KN}
 \be
\frac{{\cal B}(B\to X_s \ga)_{E_\ga> (1-\delta)E^{\rm
max}_{\ga}}}{{\cal B}(B\to X_c \ell \bar \nu)}&=& \frac{6\alpha}{\pi
f(m^2_c/m^2_b)}\frac{|V^*_{ts} V_{tb}|^2}{|V_{cb}|^2} K_{\rm
NLO}(\delta)\,,\non\\
K_{\rm NLO}(\delta)&=& \sum_{\tiny\begin{array}{c}
                       i,j=2,7\ga,8G \\
                       i\le j
                     \end{array}
}k_{ij}(\delta) {\rm Re}\left(C_i C^*_j \right) +
k^{(1)}_{77}(\delta) {\rm Re}\left( C^{(1)}_{7\ga}
C^*_{7\ga}\right)\,,
 \ed
where $\delta$ denotes the fraction of the spectrum above the cut,
$E^{\rm max}_{\ga}=m_b/2$, $f(z)=1-8z+8z^3-z^4-12z^2\ln z$ is a
phase-space factor, $K_{\rm NLO}$ stands for the next-leading-order
(NLO) effect, $C^{(1)}_{7\ga}$ is the NLO effect of $C_{7\ga}$ and
the values of $k_{ij}$ and $k^{(1)}_{77}$ are given in
Table~\ref{tab:kij}. Here, we have only considered the case with
$\delta=0.3$.
\begin{table}[hptb]
\caption{Values of $k_{ij}$ (in units of $10^{-2}$) with
$\delta=0.3$ \cite{KN} }\label{tab:kij}
\begin{ruledtabular}
\begin{tabular}{cccccccc}
$\delta$ & $k_{22}$ & $k_{77}$ & $k_{88}$ & $k_{27}$ & $k_{28}$ &
$k_{78}$ & $k^{(1)}_{77}$
 \\ \hline
0.30 & $0.11$ & $68.13$ & $0.53$ & $-16.55$ & $-0.01$ & $8.85$
&$3.86$
%
%
\end{tabular}
\end{ruledtabular}
\end{table}
According to the results in Ref.~\cite{KN}, the relevant Wilson
coefficients with charged Higgs contributions are found to be
 \be
C_{7\ga}&\approx& -0.31 +0.67 C^{\CH}_{7\ga} + 0.09 C^{\CH}_{8G}\,,\non\\
C_{8G}&\approx& -0.15 +0.70 C^{\CH}_{8G}\,,\non\\
C^{(1)}_{7\ga}&=& 0.48-2.29 C^{\CH}_{7\ga} -0.12 C^{\CH}_{8G}\,.
 \ed
We can also investigate the direct CPA for $B\to X_s \ga$,
given by \cite{KN}
 \be
A_{CP}(b\to s\ga)&=& \frac{{\cal B}(\bar B\to X_s \ga)-{\cal B}(B\to
X_s \ga)}
{{\cal B}(\bar B\to X_s \ga)+{\cal B}(B\to X_s \ga)} \,, \non\\
 &=& \frac{1}{|C_{7\ga}|^2} \left[ 1.23{\rm
Im}\left(C_2 C^{*}_{7\ga}\right)-9.52 {\rm Im}\left(C_{8G}
C^{*}_{7\ga}\right)+0.10 {\rm Im}\left(C_2 C^{*}_{8G} \right)
\right]\,, \label{eq:acp}
 \ed
where the current data is $A_{CP}(b\to s\ga)=0.004\pm 0.037$
\cite{HFAG}. Since the SM prediction is less than $1\%$
\cite{Soares}, the formula in Eq.~(\ref{eq:acp}) has neglected the
contributions related to the KM phase. If any sizable CPA is found,
it definitely indicates the existence of some new CP violating
phases. For $B\to X_s\ga$,  we first display $\phi_{q}$ versus
$\Delta m_{q}$ in Fig.~\ref{fig:phi-mq-bsga}. From the figure, it is
clear that the BR of $B\to X_s \ga$ has a very serious constraint on
$|C_{tb}|$ and $m_{\CH}$ so that the contributions of the private
charged Higgs to the time-dependent CPA become very small.
To further understand the effects of the charged Higgs on the
radiative B decays, we show the correlation between ${\cal B}(B\to
X_s\ga)$ [$A_{CP}(b\to s\ga)$] and $|C_{tb}|/m_{\CH}$ in
Figs.~\ref{fig:ctb-mh}(a)[(b)]. Interestingly, those values of
parameters, which are satisfied with the bound of ${\cal B}(B\to X_s
\ga)$, could still make $A_{CP}(b\to s\ga)$ at few percent level
where the sensitivity is the same as the current data.

\begin{figure}[bpth]
\includegraphics*[width=4 in]{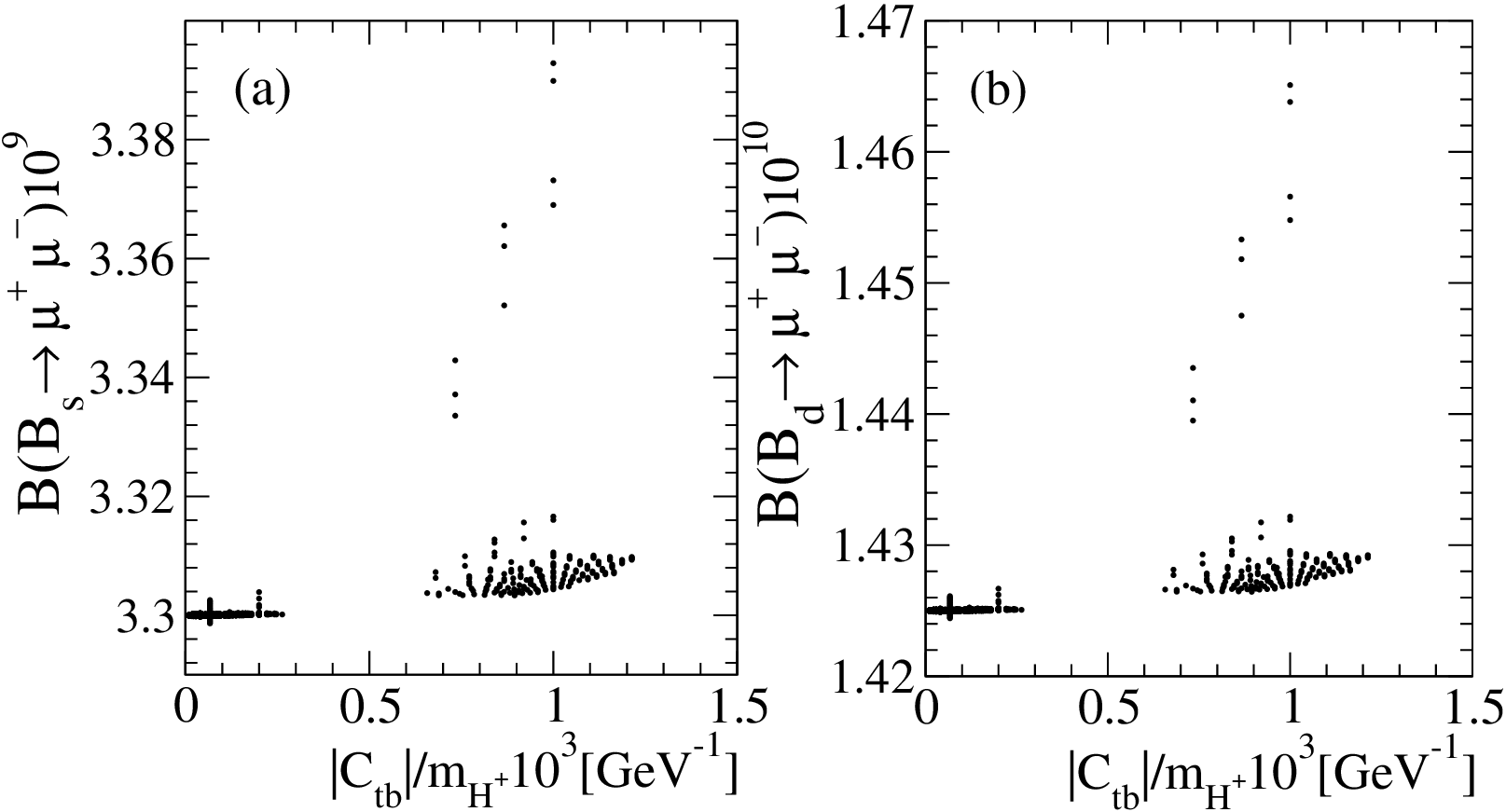}
\caption{BR of $B_{q}\to \mu^{+} \mu^{-}$ when the constraints of
$B\to X_s \ga$ are included.}
 \label{fig:Bll}
\end{figure}

\begin{figure}[bpth]
\includegraphics*[width=4 in]{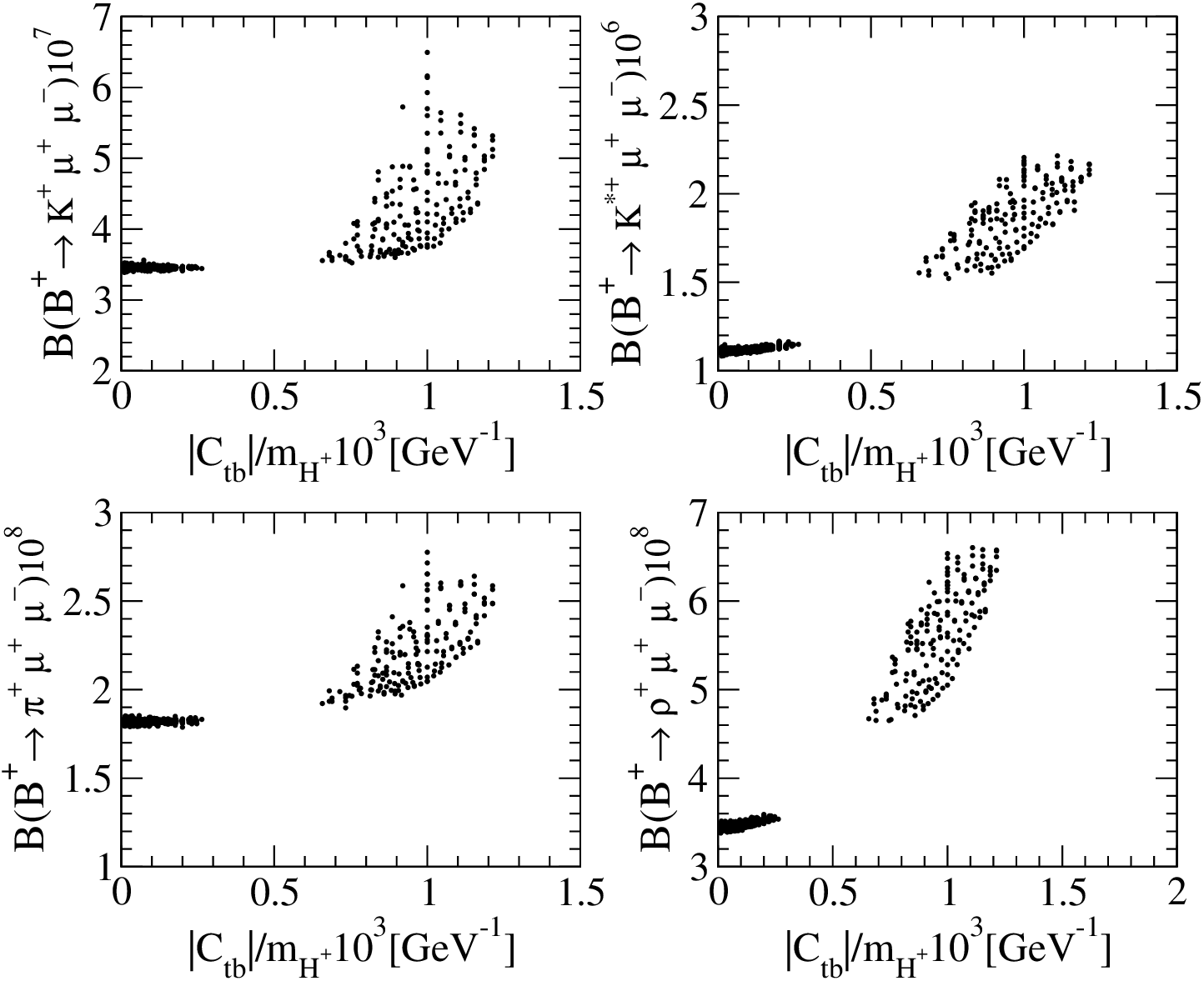}
\caption{Correlations of BRs and $|C_{tb}|/m_{\CH}$ for $B^+\to
\left( K^+,\, K^{*+},\, \pi^+,\, \rho^{+} \right)\mu^{+} \mu^{-}$.}
 \label{fig:brbMll}
\end{figure}

\begin{figure}[bpth]
\includegraphics*[width=4 in]{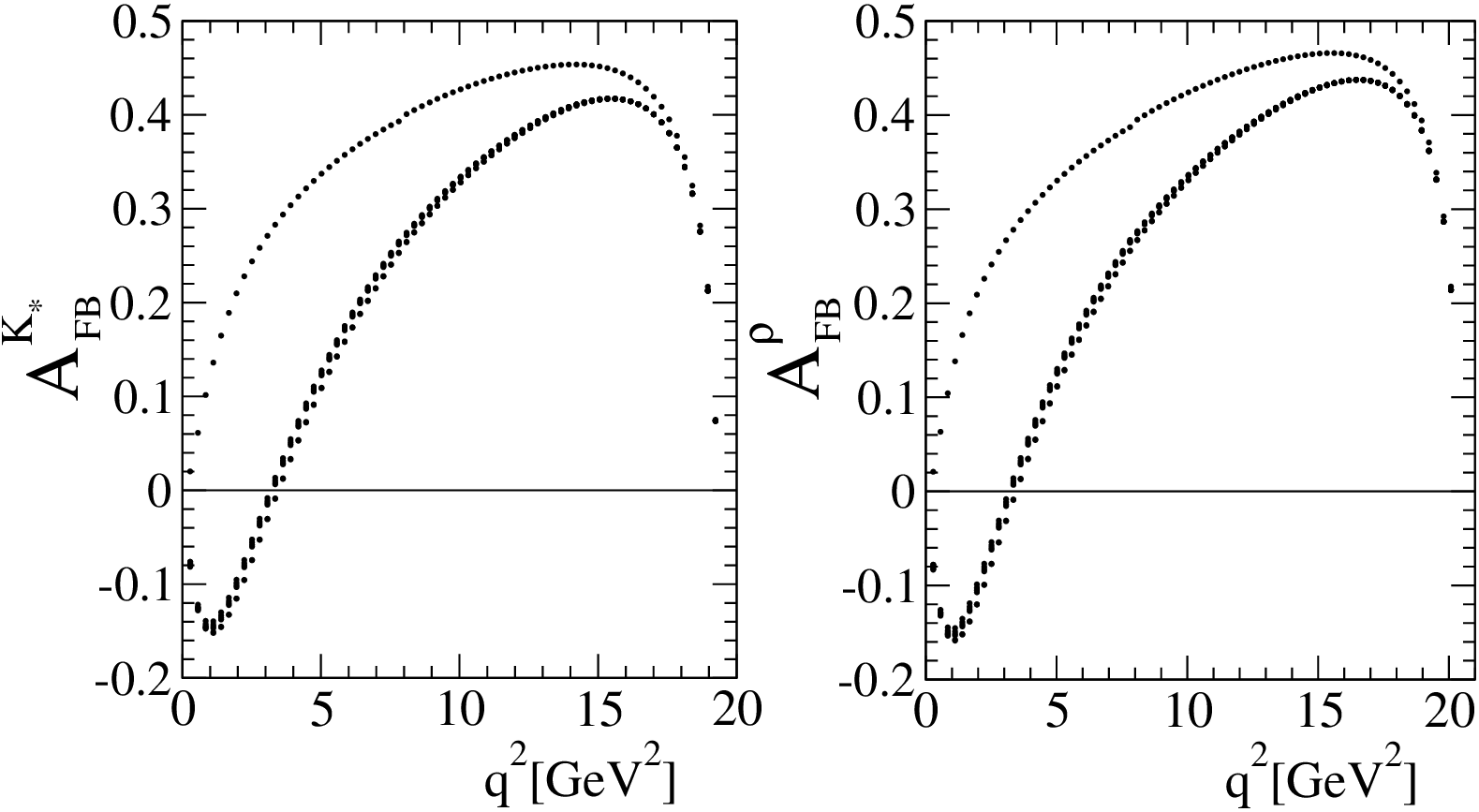}
\caption{FBAs for (a) $B^+\to K^{*+} \mu^{+} \mu^{-}$ and (b)
$B^+\to \rho^{+} \mu^{+} \mu^{-}$. } \label{fig:FBA}
\end{figure}

Next, we study the implications of the private charged Higgs on
$b\to q \ell^{+} \ell^{-}$. According to the previous analysis, we
learn that ${\cal B}(B\to X_s \ga)$ and $A_{CP}(b\to s\ga)$ could
give strong bounds on the free parameters in the PH model. With the
constraints, we show the BRs for $B_{s,d}\to \mu^{+} \mu^{-}$ in
Figs.~\ref{fig:Bll}(a) and (b). From the figures, we see that the
contributions in the PH model are very close to ${\cal
B}^{SM}(B_{s(d)}\to \mu^{+} \mu^{-})=3.3(0.14)\times 10^{-9}$ in the
SM. Hence, we conclude that the effects of Z-penguin in
Fig.~\ref{fig:bqll} are negligible.
To estimate the numerical values for $B\to (P,\, V)\ell^{+}
\ell^{-}$ decays, we use the form factors calculated by the light
cone sum rules (LCSRs), parametrized by \cite{LCSR}
 \be
 F(q^2)&=&\frac{r_{1}}{1-q^2/m^2_{1}} + \frac{r_2}{(1-q^2/m^2_2)^{n}}
 \ed
with the associated values of parameters  given in
Table~\ref{tab:PPFF} and \ref{tab:PVFF} for $B\to P$ and $B\to V$,
respectively.
\begin{table}[hptb]
\caption{ Values of parameters for $B\to (K,\, \pi)$ form factors by
LCSRs \cite{LCSR} }\label{tab:PPFF}
\begin{ruledtabular}
\begin{tabular}{cccccc}
$F(q^2)$  & $r_1$ & $m^2_1$ GeV$^2$ & $r_2$ & $m^2_2$ GeV$^2$ & $n$
 \\ \hline
$f^{B\to K}_{+}$ & $0.1616$ & $ 29.3$ & $0.173$ & $29.3$ & $1$ \\
$f^{B\to K}_{0}$ & $-$ & $ -$ & $0.3302$ & $37.46$ & $1$ \\
$f^{B\to K}_{T}$ & $0.1614$ & $ 29.3$ & $0.1981$ & $29.3$ & $1$
\\\hline
$f^{B\to \pi}_{+}$ & $0.744$ & $ 28.3$ & $-0.486$ & $40.73$ & $1$ \\
$f^{B\to \pi}_{0}$ & $-$ & $ -$ & $0.258$ & $33.81$ & $1$ \\
$f^{B\to \pi}_{T}$ & $1.387$ & $ 28.3$ & $-1.134$ & $32.22$ & $1$

\end{tabular}
\end{ruledtabular}
\end{table}
\begin{table}[hptb]
\caption{ Values of parameters for $B\to K^*(\rho)$ form factors by
LCSRs \cite{LCSR}. }\label{tab:PVFF}
\begin{ruledtabular}
\begin{tabular}{cccccc}
$F(q^2)$  & $r_1$ & $m^2_1$ GeV$^2$ & $r_2$ & $m^2_2$ GeV$^2$ & $n$
 \\ \hline
$V^{B\to K^*(\rho)}$ & $0.923(1.045)$ & $ 28.3$ & $-0.511(-0.721)$ & $49.4(38.34)$ & $1$ \\
$A^{B\to K^*(\rho)}_{0}$ & $1.364(1.527)$ & $ 28.3$ & $-0.99(-1.22)$ & $36.78(33.36)$ & $1$ \\
$A^{B\to K^*(\rho)}_{1}$ & $-$ & $ -$ & $0.29(0.24)$ & $40.38(37.51)$ & $1$ \\
$A^{B\to K^*(\rho)}_{2}$ & $-0.084(0.009)$ & $ 52(40.82)$ & $0.342(0.212)$ & $52(40.82)$ & $2$ \\
$T^{B\to K^*(\rho)}_{1}$ & $0.823(0.897)$ & $ 28.3$ & $-0.491(-0.629)$ & $46.31(38.04)$ & $1$ \\
$T^{B\to K^*(\rho)}_{2}$ & $-$ & $ -$ & $0.333(0.267)$ &
$41.41(38.59)$ &
$1$ \\
$T^{B\to K^*(\rho)}_{2}$ & $-0.036(0.022)$ & $ 48.1(40.88)$ &
$0.368(0.246)$ & $48.1(40.88)$ & $2$
\end{tabular}
\end{ruledtabular}
\end{table}
 From Eqs.~(\ref{eq:difk}) and (\ref{eq:difks}) and
with the same values of parameters for $B_q\to \mu^{+} \mu^{-}$, we
present the influence of the private charged Higgs on $B^+ \to
(K^+,\, K^{*+},\, \pi^+,\, \rho^{+})\mu^{+} \mu^{-}$ in
Fig.~\ref{fig:brbMll}. We find that the charged Higgs in the PH
model has  significant effects on the BRs for $B\to (P,\, V)\ell^{+}
\ell^{-}$. Since the contributions from $\CH$ mediated Z-penguin are
very small, the main enhancements come from the $\ga$-penguin
appearing in $C^{\CH}_{7\ga}$ of Eq.~(\ref{eq:c7ga}) and $Y^{\ga}$
of Eq.~(\ref{eq:XXY}).
By comparing with the current experimental data, expressed by
\cite{PDG08,HFAG}
 \be
{\cal B}(B^+ \to K^+ \mu^+ \mu^-)&=& (4.5^{+0.9}_{-0.8})\times 10^{-7}\,,\non\\
{\cal B}(B^+ \to K^{*+} \mu^+ \mu^-)&=&( 0.8^{+0.6}_{-0.4})\times 10^{-6} \,,\non\\
{\cal B}(B^{+} \to \pi^+ \mu^+ \mu^-)&<& 5.1\times 10^{-8}\,,
%
 \ed
we find that the BR of $B^+\to K^{*+} \mu^{+} \mu^{-}$ in the PH
model could be larger than the upper value of the current data with
$1\sigma$ error. In other words, $B^+\to K^{*+} \mu^{+} \mu^{-}$
provides a more strict constraint than $B\to X_s \ga$ does. We
notice that this result relies on the theoretical uncertainty of the
nonperturbative $B\to (P,\, V)$ form factors. However,  the QCD
errors could be controlled well with the form factors extracted from
the improved measurements on $B\to K^* \ga$ and $B\to (P,\, V) \ell
\nu$ as well as refined lattice calculations. In addition, by a more
precise measurement on $B\to K^* \ell^{+} \ell^{-}$, it is also help
to make our conclusion  more solid. Hence,  the FCNC process of
$B\to K^* \ell^{+} \ell^{-}$ has become an important candidate to
constrain the new physics. Finally, by using Eq.~(\ref{eq:FBA}), we
plot the results of the FBA in Fig.~\ref{fig:FBA}. It is clear that
the shape of the FBA for $B^+\to \rho^+  \mu^{+} \mu^{-}$ is the
same as that for $B^+\to K^{*+}\mu^+ \mu^-$ in the PH model. From
the figure, we see that there are two types of curves. The curves
crossing the zero point denote the SM-like results in which
$C_{7\ga}$ and $C^{W}_{7\ga}$ are the same sign. However, for
another type of curves, $C_{7\ga}$ and $C^{W}_{7\ga}$ are opposite
in sign. Therefore, to observe the FBA in $B\to K^*\mu^+ \mu^-$, one
can easily judge if the observed $C_{7\ga}$ has the same sign as
that in the SM.

\section{Summary}\label{sec:summary}

We have studied the charged Higgs effects in the PH model, in which
each right-handed quark is associated with one Higgs doublet in the
Yukawa sector and the hierarchy of quark masses has been represented
by the hierarchy of  the Higgs VEVs. It is found that the couplings
of the charged Higgs scalars to the fermions are independent of the
masses of quarks and order of unity when the CKM matrix elements are
excluded. Due to  $M_{b}< M_{c} \ll M_{s} \ll M_{d, u}$ of the
charged Higgs masses, we have explored the interesting effects of
these scalars in $B$ physics. By considering the constraint from the
decay of  $B\to X_s \ga$, the influence of the private charged Higgs
on the $B_q$ oscillation is negligibly small. Nevertheless, the CPA
of $B\to X_S \ga$ could reach the sensitivity of the current data.
Moreover, we have found that the BRs of $B\to (P,\, V)\ell^{+}
\ell^{-}$ are sensitive to the charged Higgs effects. With the form
factors calculated by LCSRs, we have displayed that the constraint
from the BR of $B^+\to K^{*+} \ell^{+} \ell^{-}$  could be more
stringent than that from $B\to X_s \ga$. In addition, we have shown
that the sign of $C_{7}$ in the PH model could be different from the
SM and can be further
determined by the FBA of $B\to V\ell^{+} \ell^{-}$. \\

\noindent {\bf Acknowledgments}

This work is supported in part by
the National Science Council of R.O.C. under Grant Nos:
NSC-97-2112-M-006-001-MY3 and NSC-95-2112-M-007-059-MY3. R.B is
supported by National Cheng Kung University Grant No. HUA
97-03-02-063.



\begin{thebibliography}{99}

\bibitem{PDG08}
Particle Data Group, C. Amsler {\it et al.}, Phys. Lett. B{\bf 667},
1 (2008).

\bibitem{CKM} N. Cabibbo, Phys. Rev. Lett. {\bf 10}, 531 (1963);
M. Kobayashi and T. Maskawa, Prog. Theor. Phys. {\bf 49}, 652
(1973).

\bibitem{Wolfenstein} L. Wolfenstein, Phys. Rev. Lett. {\bf 51}, 1945 (1983).

\bibitem{PZ1} R.~A.~Porto and A.~Zee, Phys. Lett. B{\bf 666}, 491 (2008)
[arXive:0712.0448 [hep-ph]].

\bibitem{PZ2} R.~A.~Porto and A.~Zee, arXiv:0807.0612.

\bibitem{Jackson} C.~B. Jackson, arXiv:0804.3792 [hep-ph].

\bibitem{Weinberg} S.~Weinberg, Phys. Rev. Lett. {\bf 37}, 657
(1976).

\bibitem{Lee} T.~D.~Lee, Phys. Rev. D{\bf 8}, 1226 (1973).

\bibitem{Masiero}  F. Gabbiani {\it et al.},
Nucl. Phys. B{\bf 477}, 321 (1996) [arXiv:hep-ph/9604387].



\bibitem{BBL}  G. Buchalla, A. J. Buras, and M. E. Lautenbacher, Rev.
Mod. Phys {\bf 68}, 1230 (1996).

\bibitem{CG_NPB636} C.~H. Chen and C.~Q. Geng, Nucl. Phys. B{\bf 636}, 338
(2002);
Phys. Rev. D{\bf 66}, 094018 (2002).

\bibitem{CG_PRD66} C.~H. Chen and C.~Q. Geng, Phys. Rev. D{\bf 66}, 014007
(2002).


\bibitem{HFAG}
  E.~Barberio {\it et al.}  [Heavy Flavor Averaging Group (HFAG)
                  Collaboration],
  arXiv:0704.3575 [hep-ex], online update at
http://www.slac.stanford.edu/xorg/hfag.


\bibitem{CDF}A. Abulencia {\it et al.} (CDF Collaboration),
Phys. Rev. Lett. {\bf 97}, 242003 (2006).

\bibitem{D0} D$\O$ note 5474-conf
at http://www-d0.fnal.gov/Run2Physics/WWW/results/prelim/B/B51.

\bibitem{bsgaSM} K. Chetyrkin {\it et al.}, Phys. Lett. B{\bf 400}, 206 (1997);
Erratum-{\it ibid.}, Phys. Lett. B{\bf 425}, 414 (1998); A.J. Buras
{\it et al.}, Phys. Lett. B{\bf 414}, 157 (1997); Erratum-{\it
ibid.}, Phys. Lett. B{\bf 434}, 459 (1998); A.~L. Kagan and M.
Neubert, Eur. Phys. J. C{\bf 7}, 5 (1999).

\bibitem{KN}A.~L. Kagan and M. Neubert, Phys. Rev. D{\bf 58}, 094012
(1998).

\bibitem{Soares}J.~M. Soares, Nucl. Phys. B{\bf 367}, 575 (1991).


\bibitem{LCSR}P. Ball and R. Zwicky, Phys. Rev. D{\bf 71}, 014015 (2005);
Phys. Rev. D{\bf 71}, 014029 (2005).

\end{thebibliography}
\end{document}